\documentclass[apj]{emulateapj}
\usepackage{mathrsfs}
\usepackage{amsfonts}
\usepackage{amssymb}
\usepackage{graphicx}
\usepackage{amsmath}
\usepackage{natbib}
\usepackage{epsfig}
\usepackage{epstopdf}
\newcommand{\asec}{\mbox{$^{\prime \prime}$}}

\def\chandra{{\it Chandra\/}}

\def\xmm{{\it XMM-Newton}}

\slugcomment{version: Aug 19th, 2012. revised for ApJ referee}
\shorttitle{Quasars SEDs in XMM-COSMOS} \shortauthors{Elvis et al.}

\begin{document}

\title{Spectral Energy Distributions of Type 1 AGN in the COSMOS Survey I - The
XMM-COSMOS Sample}

\author{M. Elvis\altaffilmark{1},
 H. Hao\altaffilmark{1, 2},
 F. Civano\altaffilmark{1},
 M. Brusa\altaffilmark{3},
 M. Salvato\altaffilmark{3,4,5},
 A. Bongiorno\altaffilmark{3,6},
 P. Capak\altaffilmark{7},
 G. Zamorani\altaffilmark{8},
 A. Comastri\altaffilmark{8},
 K. Jahnke\altaffilmark{9},
 E. Lusso\altaffilmark{9},
 V. Mainieri\altaffilmark{10},
 J. R. Trump\altaffilmark{11,12},
 L. C. Ho\altaffilmark{13},
H.~Aussel\altaffilmark{14}, N. Cappelluti\altaffilmark{3,8},
M.~Cisternas\altaffilmark{9}, D. Frayer\altaffilmark{15},
R.~Gilli\altaffilmark{8}, G. Hasinger\altaffilmark{16}, J.
P.~Huchra\altaffilmark{1,17}, C. D. Impey\altaffilmark{11}, A.
M.~Koekemoer\altaffilmark{18}, G. Lanzuisi\altaffilmark{1,3,19, 20},
E. Le Floc'h\altaffilmark{21}, S.~J. Lilly\altaffilmark{22},
Y.~Liu\altaffilmark{23}, P. McCarthy\altaffilmark{13},
H.~J.~McCracken\altaffilmark{24}, A. Merloni\altaffilmark{3},
H.-J.~Roeser\altaffilmark{9}, D. B. Sanders\altaffilmark{16},
M.~Sargent\altaffilmark{9, 21}, N. Scoville\altaffilmark{7},
E.~Schinnerer\altaffilmark{9}, D. Schiminovich\altaffilmark{25},
J.~Silverman\altaffilmark{26}, Y. Taniguchi\altaffilmark{27},
C.~Vignali\altaffilmark{28}, C. M. Urry\altaffilmark{29}, M.
A.~Zamojski\altaffilmark{25}, M. Zatloukal\altaffilmark{9}}

\altaffiltext{1}{Harvard Smithsonian Center for astrophysics, 60
Garden St., Cambridge, MA 02138, USA}

\altaffiltext{2}{SISSA, Via Bonomea 265, I-34136 Trieste, Italy}

\altaffiltext{3}{Max Planck Institute f\"ur Extraterrestrische
Physik, Postfach 1312, 85741, Garching bei M\"{u}nchen, Germany}

\altaffiltext{4}{IPP - Max-Planck-Institute for Plasma Physics,
Boltzmann Strasse 2, D-85748, Garching bei M\"{u}nchen, Germany}

\altaffiltext{5}{Excellence Cluster, Boltzmann Strasse 2, D-85748,
Garching bei M\"{u}nchen, Germany}

\altaffiltext{6}{INAF-Osservatorio Astronomico di Roma, Via di
Frascati 33, 00040, Monteporzio Catone, Rome, Italy}

\altaffiltext{7}{California Institute of Technology, MC 105-24, 1200
East California Boulevard, Pasadena, CA 91125, USA}

\altaffiltext{8}{INAF-Osservatorio Astronomico di Bologna, via
Ranzani 1, I-40127 Bologna, Italy}

\altaffiltext{9}{Max Planck Institute f\"ur Astronomie, K\"onigstuhl
17, Heidelberg, D-69117, Germany}

\altaffiltext{10}{European Southern Observatory,
Karl-Schwarzschild-Strasse 2, D-85748 Garching bei M\"{u}nchen,
Germany}

\altaffiltext{11}{Steward Observatory, University of Arizona, 933
North Cherry Avenue, Tucson, AZ 85721, USA}

\altaffiltext{12}{UCO/Lick Observatory, University of California,
Santa Cruz, CA 95064, USA}

\altaffiltext{13}{The Observatories of the Carnegie Institute for
Science, Santa Barbara Street, Pasadena, CA 91101, USA}

\altaffiltext{14}{AIM Unit\'e Mixte de Recherche CEA CNRS,
Universit\'e Paris VII UMR n158, Paris, France}

\altaffiltext{15}{National Radio Astronomy Observatory, P.O. Box 2,
Green Bank, WV 24944, USA}

\altaffiltext{16}{Institute for Astronomy, University of Hawaii,
2680 Woodlawn Drive, Honolulu, HI 96822, USA}

\altaffiltext{17}{John P. Huchra has contributed to the work before
his death in October 2010.}

\altaffiltext{18}{Space Telescope Science Institute, 3700 San Martin
Drive, Baltimore, MD 21218, USA}

\altaffiltext{19}{INAF-IASF Roma, Via Fosso del Cavaliere 100, 00133
Rome, Italy}

\altaffiltext{20}{INAF-IASF Bologna, Via Gobetti 101, I-40129
Bologna, Italy}

\altaffiltext{21}{CEA-Saclay, Service d'Astrophysique, Orme des
Merisiers, Bat. 709, 91191 Gif-sur-Yvette, France}

\altaffiltext{22}{Institute of Astronomy, Swiss Federal Institute of
Technology (ETH H\"onggerberg), CH-8093, Z\"urich, Switzerland}

\altaffiltext{23}{Key Laboratory of Particle Astrophysics, Institute
of High Energy Physics, Chinese Academy of Sciences, P.O. Box 918-3,
Beijing 100049, China}

\altaffiltext{24}{Institut d'Astrophysique de Paris, UMR 7095 CNRS,
Universit\'e Pierre et Marie Curie, 98 bis Boulevard Arago, F-75014
Paris, France}

\altaffiltext{25}{Department of Astronomy, Columbia University,
MC2457,550 W. 120 St. New York, NY 10027, USA}

\altaffiltext{26}{Institute for the Physics and Mathematics of the
Universe (IPMU), University of Tokyo, Kashiwanoha 5-1-5,
Kashiwa-shi, Chiba 277-8568, Japan}

\altaffiltext{27}{Research Center for Space and Cosmic Evolution,
Ehime University, Bunkyo-cho 2-5, Matsuyama 790-8577, Japan}

\altaffiltext{28}{Dipartimento di Astronomia, Universit\`{a} degli
Studi di Bologna, via Ranzani 1, I-40127 Bologna, Italy}

\altaffiltext{29}{Physics Department and Yale Center for Astronomy
and Astrophysics, Yale University, New Haven, CT 06511, USA}

\email{elvis@cfa.harvard.edu, hhao@cfa.harvard.edu}

\begin{abstract}

The ``Cosmic Evolution Survey" (COSMOS) enables the study of the
Spectral Energy Distributions (SEDs) of Active Galactic Nuclei (AGN)
because of the deep coverage and rich sampling of frequencies from
X-ray to radio. Here we present a SED catalog of 413 X-ray (\xmm)
selected type~1 (emission line FWHM$>2000$~km~s$^{-1}$) AGN with
Magellan, SDSS or VLT spectrum. The SEDs are corrected for the
Galactic extinction, for broad emission line contributions,
constrained variability, and for host galaxy contribution. We
present the mean SED and the dispersion SEDs after the above
corrections in the rest frame 1.4 GHz to 40~keV, and show examples
of the variety of SEDs encountered. In the near-infrared to optical
(rest frame $\sim 8\mu m$-- 4000\AA), the photometry is complete for
the whole sample and the mean SED is derived from detections only.
Reddening and host galaxy contamination could account for a large
fraction of the observed SED variety. The SEDs are all available
on-line.
\end{abstract}


\section{Introduction}

Quasars and Active Galactic Nuclei (AGN) are the most luminous
persistent sources of radiation in the universe. The AGN luminosity
is emitted primarily in a broad continuum spectrum that carries
significant power over several decades, from the far-infrared (FIR)
to the X-ray bands. Hence, knowing the spectral energy distributions
(SEDs) of AGN is essential to a deeper understanding of quasar
physics. The mean spectral energy distribution (SED) compiled by
Elvis et al. (1994, ``E94'' hereinafter) is still the most commonly
used SED for quasars, despite recent additions and updates (see e.g.
Polletta et al. 2000, Kuraszkiewicz et al. 2003, Marconi et al.
2004, Risaliti \& Elvis 2004, Richards et al. 2006, Hopkins et al.
2006, Polletta et al. 2007, Shang et al. 2011, Luo et al. 2010,
Lusso et al. 2010, 2011). However the E94 SEDs suffer from
significant limitations: (1) the sample is primarily an ultraviolet
excess selected sample, and is further biased toward relatively
X-ray loud quasars; (2) the mean SED was compiled from a small
number of AGN (29 radio-quiet and 18 radio-loud AGN); (3) the sample
only covers a low redshift range ($0.05\leq z \leq 0.9$, with 80\%
being at $z<0.3$); (4) the data in X-ray, ultraviolet and far
infrared region have limited signal-to-noise ratios, with a large
number of upper limits. Even so, E94 found a large dispersion in SEDs of
$\sim$1~dex at both 100~$\mu$m and 0.1~$\mu$m, when all the SEDs
were normalized at 1~$\mu$m.  The 1~$\mu$m wavelength is usually
chosen as a normalization point because it is the approximate
location of the inflection between the rising Wien tail of emission
from hot dust and the power-law $f_\nu \propto \nu^{1/3}$ of the big
blue bump in the optical caused by the emission of the accretion
disk in $\nu f_\nu$ versus $\nu$ space (E94). This variety of
continuum shapes in quasars has not been carefully explored so that
no correlations of SED shape in the FIR-UV with other properties
have been found, nor is there an accepted theoretical explanation
fitting all the various forms.

The Cosmic Evolution Survey (COSMOS; Scoville et al. 2007a) has the
appropriate combination of depth, area and multi-wavelength coverage
that allows detection of substantial AGN samples by all standard
techniques -- X-ray (Brusa et al. 2007, 2010, Civano et al. 2012),
infrared (Donley et al. 2012), radio (Schinnerer et al. 2010) and
optical (Gabor et al. 2009). COSMOS covers a 2 deg$^2$ equatorial
field centered on $RA=10:00:28.6$, $Dec=+02:12:21$ (J2000) with deep
imaging by most of the major space-based telescopes: {\it Hubble}
(Scoville et al. 2007b, Koekemorer et al. 2007), {\it Spitzer}
(Sanders et al. 2007, 2010), {\it GALEX} (Zamojski et al. 2007),
\xmm\ (Hasinger et al. 2007, Cappelluti et al. 2009), \chandra\
(Elvis et al. 2009), and large ground based telescopes: Subaru
(Taniguchi et al. 2007), VLA (Schinnerer et al. 2007, 2010), CFHT
(McCracken et al. 2007), UKIRT, NOAO (Capak et al. 2007). The
imaging coverage is complemented by dedicated redshift surveys
conducted with the VIMOS/VLT (zCOSMOS, Lilly et al. 2007, 2009),
IMACS/Magellan (Trump et al. 2007, 2009a), DEIMOS/Keck\footnote{The
result of a multi-year observing campaign (PIs: Capak, Kartaltepe,
Salvato, Sanders, Scoville; see Kartaltepe et al. 2010).}, for a
total of more than 20k spectra collected. The field has also been
covered by SDSS (Abazajian et al. 2004), for both photometry and
spectroscopy.

COSMOS contains over 400 type~1 (i.e.  broad emission line, FWHM $>$
2000 km s$^{-1}$) AGN, spanning a wide range of redshifts, that were
identified through the MMT/IMACS, zCOSMOS and SDSS surveys (Brusa et
al. 2007, 2010). So far the COSMOS data set has 43 photometric
bands, with high signal-to-noise ($\gtrsim$80) for a typical I=22.5
type~1 AGN. Hence COSMOS overcomes the limitations of E94, offering
an opportunity of making order of magnitude improvements in our
knowledge of AGN SEDs.

This is the first of a series of papers on the SEDs of the X-ray
selected type~1 AGN in COSMOS. The main goal of this study is to
improve the type~1 AGN SED template over that of E94 and, perhaps
more importantly, to study the diversity of AGN SEDs and their
relation with physical parameters. Lusso et al. (2010), in a
complementary paper, analyzed a sample of 545 \xmm\ selected type 1
AGN in XMM-COSMOS, including both the spectroscopic (361/545) and
photometric (184/545) identifications in order to study dependence
on luminosity and redshift of the relationship between the UV and
X-ray luminosity ($\alpha_{OX}$). Trump et al. (2011) also presented
SED information for 348 XMM-Newton selected AGN in COSMOS, although
they focused on the differences between type 1 and unobscured type 2
AGN SEDs rather than the dispersion among type 1 AGN themselves.
This paper (Paper~I) describes the properties of the sample used for
this study and presents the mean SED, including corrections for the
Galactic extinction, broad emission line (BEL) contribution, while
limiting both variability and host galaxy contamination. Other 3
companion papers on this subject are in preparation (Hao et al. 2012
a, b, c). The companion Paper II (Hao et al. 2012a) investigates the
systematic trends in the shapes of the SED in the wavelength range
of 0.3-3$\mu m$ as a function of redshift, bolometric luminosity,
black hole mass and Eddington ratio. Paper~III (Hao et al. 2012b)
introduces a ``mixing diagram'' (near-infrared SED slope versus
optical SED slope plot) to understand the diversity of the SED
shapes in XMM-COSMOS type 1 AGN sample, provide a new estimation of
the host galaxy fraction and reddening, and identify interesting
outliers. Paper~IV (Hao et al. 2012c) studies the radio-loudness of
the XMM-COSMOS type 1 AGN. An interesting sub-sample of ``hot dust
poor'' quasars, drawn from this sample, have already been presented
in Hao et al. (2010).

In this paper, all magnitudes are reported in the AB system (Oke \&
Gunn 1983) and the WMAP 5-year cosmology (Komatsu et al. 2009),
with H$_0$ =71~km~s$^{-1}$~Mpc$^{-1}$, $\Omega_M$ = 0.26 and
$\Omega_{\Lambda}$= 0.74 is assumed.

\section{Sample Description}
\begin{deluxetable*}{cccclcccc}
\tabletypesize{\scriptsize} \tablecaption{Data Quality and
Depth\label{t:data-depth}} \tablehead{ \colhead{Filter}  &\colhead{
Telescope} & \colhead{Effective\tablenotemark{1}} &
\colhead{Filter\tablenotemark{2}} & \colhead{Depth\tablenotemark{3}}
& \colhead{Number of } &\colhead{Observations}&
\colhead{$k_{\lambda}$\tablenotemark{4}}& \colhead{$A_{\lambda}$\tablenotemark{5}}\\
\colhead{Name} &      & \colhead{Wavelength(\AA)} & \colhead{
Width(\AA)} &  \colhead{}     &\colhead{Detections} & \colhead{date}
& & } \startdata
$X$ & XMM & 2keV&0.5-10keV&$10^{-15}$\tablenotemark{**} & 413 &Dec 2003-May 2005&&\\
$FUV$ & Galex & 1540 & 209 & 25.69 & 136  & Feb 2004 & 8.92 & 0.17\\
$NUV$ & Galex & 2315 & 797 & 25.99  & 263 & Feb 2004 & 7.97 & 0.15\\
$u$ & SDSS & 3564 & 600 & 22.00  & 370 & Jul 2001\tablenotemark{*} & 4.69&0.09\\
$u^*$ & CFHT & 3823 & 605   & 26.50  & 412  & Jan 2003-Apr 2007 & 4.69&0.09\\
$IA427$ & Subaru & 4263 & 208 & 25.82  & 411  & Jan 2006 & 4.21&0.08\\
$B_J$ & Subaru & 4458 & 897 & 27.00& 413  & Jan 2004 & 4.04&0.08 \\
$IA464$ & Subaru & 4635 & 218 & 25.65  & 411 & Feb 2006 &3.86 &0.07\\
$g$ & SDSS & 4723 & 1300 & 22.20  & 405  & Jul 2001\tablenotemark{*} & 3.74&0.07 \\
$g^+$ & Subaru & 4777 & 1265 & 27.00 & 413  & Feb 2004 & 3.74&0.07\\
$IA484$ & Subaru & 4849 & 229 & 25.60 & 410 & Jan 2007 & 3.67&0.07\\
$IA505$ & Subaru & 5063 & 232 & 25.55  & 412 & Feb 2006 & 3.45&0.07\\
$IA527$ & Subaru & 5261 & 243 & 25.62 & 410 & Jan 2007 & 3.29&0.06\\
$V_J$ & Subaru & 5478 & 946 & 26.60  & 412 & Feb 2004 & 3.15&0.06\\
$IA574$ & Subaru & 5765 & 273 & 25.61 & 412 & Jan 2006 & 2.96&0.06\\
$r$ & SDSS & 6202 & 1200 & 22.00 & 401 & Jul 2001\tablenotemark{*} & 2.59&0.05\\
$IA624$ & Subaru & 6233 & 300 & 25.60 & 410 & Dec 2006 & 2.61&0.05\\
$r^+$ & Subaru & 6289   & 1382 & 26.80 & 413 & Jan 2004 & 2.59&0.05\\
$IA679$ & Subaru & 6781 & 336 & 25.60  & 412 & Feb 2006 & 2.27&0.04\\
$IA709$ & Subaru & 7074 & 317 & 25.65  & 411 & Jan 2006 & 2.15&0.04\\
$NB711$ & Subaru & 7120 & 73 & 25.00 & 412 & Feb 2006 & 2.13&0.04\\
$IA738$ & Subaru & 7362 & 324 & 25.60 & 410 & Jan 2007 & 2.04&0.04\\
$i$ &SDSS & 7523 & 1300 & 21.30 & 406 & Jul 2001\tablenotemark{*} & 1.92&0.04\\
$i^*$    & CFHT & 7618 & 1300 & 24.00 & 411 & Jan 2004 & 1.92&0.04\\
$i^+$ &Subaru   & 7684 & 1495 & 26.20  & 199\tablenotemark{6} & Jan 2004 & 1.92&0.04\\
$IA767$ & Subaru & 7685 & 364 & 25.60 & 412 & Mar 2007 & 1.92&0.04\\
$F814W$ & HST & 8072 & 1830 & 27.10 & 388\tablenotemark{7} &Oct 2003\tablenotemark{*} & 1.80&0.03\\
$NB816$ & Subaru & 8149 & 120 & 25.70 & 413 & Feb 2005\tablenotemark{*} & 1.74&0.03\\
$IA827$ & Subaru & 8245 & 343 & 25.39 & 411 & Jan 2006 & 1.69&0.03\\
$z$ & SDSS & 8905 & 1000 & 20.50 & 368 & Jul 2001\tablenotemark{*} & 1.44&0.03\\
$z^+$ & Subaru & 9037 & 856 & 25.20 & 411 & Jan 2004 & 1.44&0.03\\
$J$ & UH 88" & 12491 & 1580 & 23.70 & 413 & Mar 2006 & 0.97&0.02\\
$H$ & Calar Alto & 16483 & 2665 & 20.90 & 252 & Aug 2005 & &\\
$K$ & KPNO & 21537 & 3120 & 21.60 & 406 & Feb 2004\tablenotemark{*} & 0.34&0.01\\
$K$ & CFHT & 21590 & 3255 & 23.70 & 413 & Mar 2007 & 0.34&0.01\\
$IRAC1$ & Spitzer & 35635 & 7430 & 23.90 & 413 & Jan 2006 & 0.34&0.01\\
$IRAC2$ & Spitzer & 45110 & 10110 & 23.30 & 413 & Jan 2006 & 0.33&0.01\\
$IRAC3$ & Spitzer & 57593 & 14060 & 21.30 & 413 & Jan 2006 & 0.33&0.01\\
$IRAC4$ & Spitzer & 79594 & 28760 & 21.00 & 413 & Jan 2006 & 0.32&0.01\\
$MIPS24$ & Spitzer & 236741 & 50560 &$80\mu Jy$ & 385 & Jan 2006-Jan 2008 &&\\
$MIPS70$ & Spitzer & 714329 & 184844 & 4 mJy &34& Jan 2006-Jan 2008 &&\\
$MIPS160$ & Spitzer & 1558938 & 344982 & 30 mJy &8& Jan 2006-Jan 2008 &&\\
$L$ & VLA &1.4GHz&75MHz&$45\mu Jy$&139\tablenotemark{8}&Aug 2003-Mar 2006&&\\
\enddata

\tablenotetext{1}{Effective wavelength: $\lambda_{eff}=\int
  R*\lambda d\lambda/\int R d\lambda$, where
  R is the transmission profile normalized to a peak throughput
   of unity, and including the
transmission of the atmosphere, the telescope, the camera optics,
the filter and the detector.} \tablenotetext{2}{The FWHM of the
response function.} \tablenotetext{3}{$5\sigma$ in a 3\asec aperture
for data from infrared to UV bands. Flux limit for the radio and
X-ray (in erg cm$^{-2}$ s$^{-1}$) data.}
\tablenotetext{4}{$k_{\lambda}$ is the constant parameter for each
  band to calculate the Galactic extinction by $A_{\lambda}=k_{\lambda}E(B-V)$.}
\tablenotetext{5}{Galactic extinction calculated by $k_{\lambda}$ in
each band multiplied by mean E(B-V) 0.0192 (see \S~\ref{s:ebvcor}).}
\tablenotetext{6}{The rest of the sources are so bright that they
saturate in the band.} \tablenotetext{7}{The rest are out of the
region of the Hubble coverage.} \tablenotetext{8}{61 sources have
$>5\sigma$ detections, 78 sources have $3\sigma \sim 5\sigma$
detections. In addition, 268 other sources have $3\sigma$ upper
limits.} \tablenotetext{*}{These photometry data were excluded after
variability correction.} \tablenotetext{**}{The unit is
erg/(cm$^2$s).}

\end{deluxetable*}

X-ray emission is ubiquitous in AGN, and in the X-ray band, the
obscuration and complication from host galaxy light are minimized.
Hence X-ray surveys give the most complete and effective census of
AGN of any single band (Risaliti \& Elvis 2004). The XMM-COSMOS
survey (Hasinger et al. 2007) observed the entire COSMOS field for a
total of $\sim 1.5$~Ms, with an average exposure of $60$~ks across
the field to a depth of $\sim 5\times10^{-16}$~erg~cm$^{-2}$s$^{-1}$
($0.5-2$~keV) and $\sim3\times10^{-15}$~erg~cm$^{-2}$s$^{-1}$
($2-10$~keV) over 90\% of the surveyed area. As such XMM-COSMOS is
the deepest survey over such a large contiguous solid angle
(2.13~deg$^2$) performed to date with XMM-Newton (Cappelluti et al.
2009). X-ray fluxes were computed from the count rates assuming the
Galactic column density of $N_H=2.6\times 10^{20} cm^{-2}$ (Dickey
\& Lockman 1990), and spectral indices of $\Gamma=2$ for the
$0.5-2$~keV band, and $\Gamma=1.7$ for the $2-10$~keV band
(Cappelluti et al. 2009).  A total of 1848 point-like sources were
detected: 1567 in the $0.5-2$~keV, 1096 in the $2-8$~keV band, and
245 in the $4.5-10$~keV band, excluding those associated with
extended X-ray sources.

Brusa et al. (2007, 2010) used the likelihood ratio technique
(Sutherland \& Saunders 1992, Ciliegi et al. 2003, Brusa et al.
2005, Civano et al. 2012) to identify the counterparts of XMM-COSMOS
sources using the optical (CFHT I band), near-infrared (CFHT K band)
and mid-infrared (IRAC) catalogs. For the subfield of the XMM–COSMOS
covered also by Chandra, the optical identifications have been
augmented with the more accurate Chandra positions (Civano et al.
2012). 85\% (1577) of the \xmm\ sources have a unique and secure
optical counterpart with a probability of misidentification of
$<0.01$ (i.e. no ambiguous optical counterparts, no \xmm\ sources
separated into two sources in the \chandra\ catalog, see Table 1 of
Brusa et al. 2010 for details.). The counterparts have been
cross-correlated with the full COSMOS photometric catalog (Capak et
al. 2007), including optical bands from CFHT, SDSS and Subaru
multiwavelength bands (Taniguchi et al. 2007), which consists of 6
broad ($\Delta\lambda=$800--1500~\AA) bands, 12 intermediate
($\Delta\lambda=$300--500~\AA) bands and 2 narrow
($\Delta\lambda=$150--200~\AA) bands. The counterparts have also
been cross-correlated with the CFHT J-band (Ilbert et al. 2009),
CFHT K-band (Mc Cracken et al. 2010) catalog. The sources were also
associated with {\it GALEX} sources from the deblended, PSF-fitted
{\it GALEX} COSMOS catalog (Zamojski et al 2007), and with the IRAC
(Sanders et al. 2007), the MIPS 24$\mu m$ (Le Floc'h et al. 2009),
and the 70$\mu m$ (Frayer et al. 2009, Kartaltepe et al. 2010)
catalogs. More details on cross-correlation procedure and source
identification are available in Brusa et al. (2010).

The counterpart positions were then cross-correlated with all the
available redshift catalogs in the field, including the most recent
SDSS release (Schneider et al. 2007), the COSMOS AGN spectroscopic
survey and the zCOSMOS survey catalogs. The Magellan/IMACS AGN
spectroscopic survey has optical spectra of an X-ray and optical
flux-limited sample (down to $f_{0.5-10~keV}> 8\times 10^{-16}$
erg~cm$^{-2}$s$^{-1}$ and $i^{+}_{AB}<23$ ) of 677 XMM-Newton
selected sources over the entire COSMOS field. Trump et al. (2009a)
find 485 high-confidence redshifts from the first 3 years of the
Magellan/IMACS survey, with a total of 588 AGN with high-confidence
redshifts at the survey's completion. zCOSMOS is a redshift survey
of 20,000 galaxies in the COSMOS field taken with the VLT/VIMOS
spectrograph. The zCOSMOS bright sample is magnitude-limited with
$i^{+}_{AB}<22.5$. \xmm\ source counterparts were targeted
explicitly. About 500 \xmm\ sources have zCOSMOS spectra (many of
which also have IMACS spectra). In summary, a total of 886 unique
good quality spectroscopic redshifts are available, or $\sim$ 50\%
of the entire XMM-COSMOS sample (see details in Brusa et al. 2010).
From this sample, we selected all the objects identified as type 1
AGN (i.e. those showing broad emission lines with FWHM$>$2000 km
$s^{-1}$ in their optical spectra). This gives the final sample of
413 type 1 XMM-COSMOS AGN: the ``XC413'' sample.

The near-infrared data set of COSMOS so far has only the J and K
bands, leading to rather sparse coverage of the rest frame optical
parts of the SEDs for redshifts $\sim$1--2.  H-band imaging of the
COSMOS field to $H_{AB}\sim 21$ has been obtained as part of the
Heidelberg InfraRed/Optical Cluster Survey (HIROCS survey; Falter et
al. 2004). H-band total magnitudes (obtained through PSF fitting)
for 218 type 1 AGN in this sample were supplied to us (H.-J.
R\"{o}ser, private communication).

1.4 GHz counterparts to the X-ray sources were determined by
positional matching the optical coordinates of the X-ray sources to
the positions in the VLA-COSMOS Joint catalog (Schinnerer et al.
2010) with a search radius of 1\asec. The VLA-COSMOS joint catalog
lists ~2900 sources detected at S/N$\geqslant$5 in the COSMOS field.
All successful matches (61 sources) were unique. The AIPS/MAXFIT
peak finding algorithm was used to search for additional radio
detections within a 2.5\asec$\times$2.5\asec\ box centered on the
optical coordinates of unmatched X-ray sources. For the 78
detections in the range 3-5 $\sigma$, we computed their total flux
using the assumption that they are not resolved at 1.4GHz (beam FWHM
2.5\asec). For flux peaks with a lower significance level we derived
a 3$\sigma$ upper limit based on the local rms noise (calculated
within a 17.5\asec$\times$17.5\asec\ box) at the position of the
radio source.

To derive SEDs for the XC413, we made use of all the data described
above.  Overall, the photometry is complete in 30 bands for more
than $\sim$99\% of the XC413 sample, and it is complete in 5
additional bands for more than $\sim$90\% of the sample. In the
infrared to ultraviolet wavelength range, where the bulk of the AGN
power is radiated, we have complete coverage in the 6 infrared bands
(IRAC 1--4, J and K), and complete, richer, coverage in the optical
(Subaru broad and intermediate bands). This makes the XC413 sample
the best so far with which to study the infrared to optical SED in
great detail. Table 1 reports the bands used, their effective
wavelengths and filter widths, the limiting depth of the survey, the
number of XC413 sources having a detection in each band, and the
observation dates. For convenience we also tabulate the value of
A$_{\lambda}$ for the typical Galactic extinction and the
coefficient, $k_{\lambda}$, used for each band to calculate the
Galactic extinction.

\section{Sample Properties}

\begin{figure}
\epsfig{file=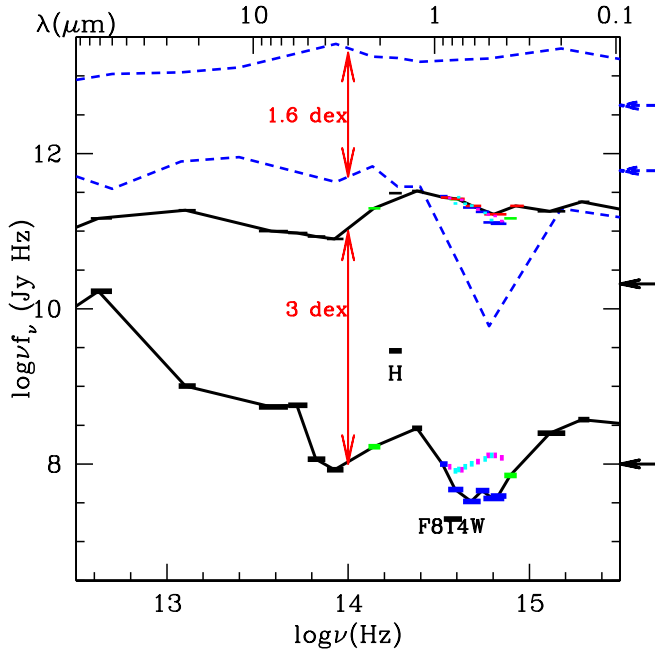, angle=0,width=\linewidth} \caption{Spectral
window function of the 413 type 1 AGN in XMM-COSMOS (XC413, black
solid line) compared to the E94 (blue dashed line) sample.  The
spectral window function consists of two lines. The upper line shows
the brightest data and the lower line shows the limits of each band.
From the left to right the points are: MIPS70, MIPS24, IRAC4 to
IRAC1 (black), CFHT K band (green), H band (black), J band (black),
ACS F814W band (black), Subaru broad bands (blue), SDSS (red),
Subaru intermediate bands (2006 magenta, 2007 cyan), CFHT u band
(green), GALEX NUV and FUV bands (black). The X-ray window width at
2 keV is shown as arrows on the right. \label{fwin}}
\end{figure}

The depth of the multi-wavelength coverage in COSMOS results in a
wide range of detectable fluxes at most wavelengths. In Figure~
\ref{fwin}, we show the COSMOS ``spectral window function''
(Brissenden 1989) over the infrared to ultraviolet bands of XC413 in
log$\nu f_{\nu}$ versus log$\nu$ space. The spectral window function
is composed of two curves: the upper one picks out the brightest
objects in the XC413 sample in each band; the lower one shows the
limiting flux for detection in each band. The spectral window
function thus shows the range of SEDs detectable for a sample. The
width of the window function in the X-ray band is also indicated
with arrows on the right. The XC413 AGN spectral window function has
a typical width of $\sim$3~dex in the infrared--optical--ultraviolet
part of the spectrum where the bulk of the AGN power is radiated.
This is about twice the range in the E94 sample (dashed blue lines)
in almost all of the bands. Hence, the XC413 sample is capable of
revealing a larger variety of type~1 AGN SEDs than the E94 sample
(see \S~\ref{s:variety} for details).

\subsection{Luminosity and Redshift}

\begin{figure}
\epsfig{file=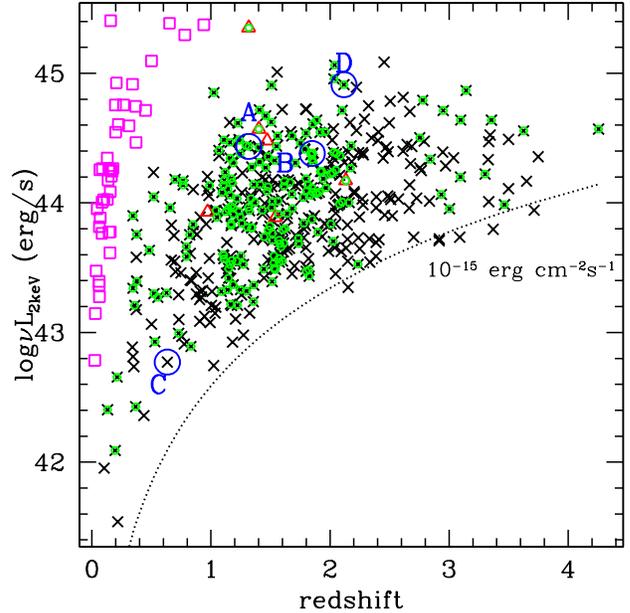, angle=0,width=\linewidth} \caption{X-ray
luminosity versus redshift for the 413 type 1 AGN in XMM-COSMOS. The
depth of the survey is $\sim5\times10^{-16}$~erg~cm$^{-2}$s$^{-1}$
($0.5-2$~keV) and $\sim3\times10^{-15}$~erg~cm$^{-2}$s$^{-1}$
($2-10$~keV) over 90\% of the area. The dotted line shows the X-ray
flux of $10^{-15}$~erg~cm$^{-2}$s$^{-1}$ as a guidance of the flux
limit. Quasars from E94 are shown as magenta squares. Black crosses
= radio quiet sources, red triangles = radio loud sources (see
\S~\ref{s:radio} for the definition). Green hexagons = sources
corrected for host galaxy contribution (see \S~\ref{s:host}). The
four sources labeled (A, B, C, D) in blue are discussed in
\S~\ref{s:variety}.
  \label{Lxz}}
\end{figure}

\begin{figure}
\epsfig{file=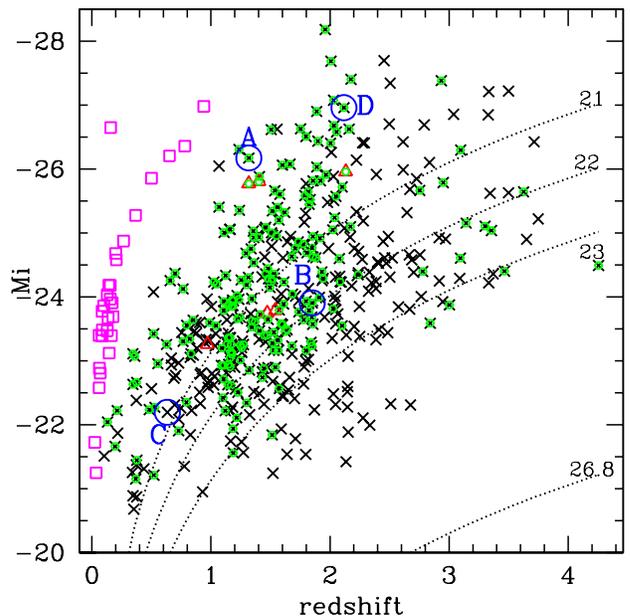, angle=0,width=\linewidth} \caption{I band
absolute magnitude versus redshift for the 413 type 1 AGN in
XMM-COSMOS. The dotted lines represent the SDSS photometry flux
limit (i=21 mag), the Magellan and MMT flux limit (i=23 mag) and the
photometry flux limit (r=26.8 mag). The symbols are as in
Figure~\ref{Lxz}.\label{mz}}
\end{figure}

\begin{figure}
\epsfig{file=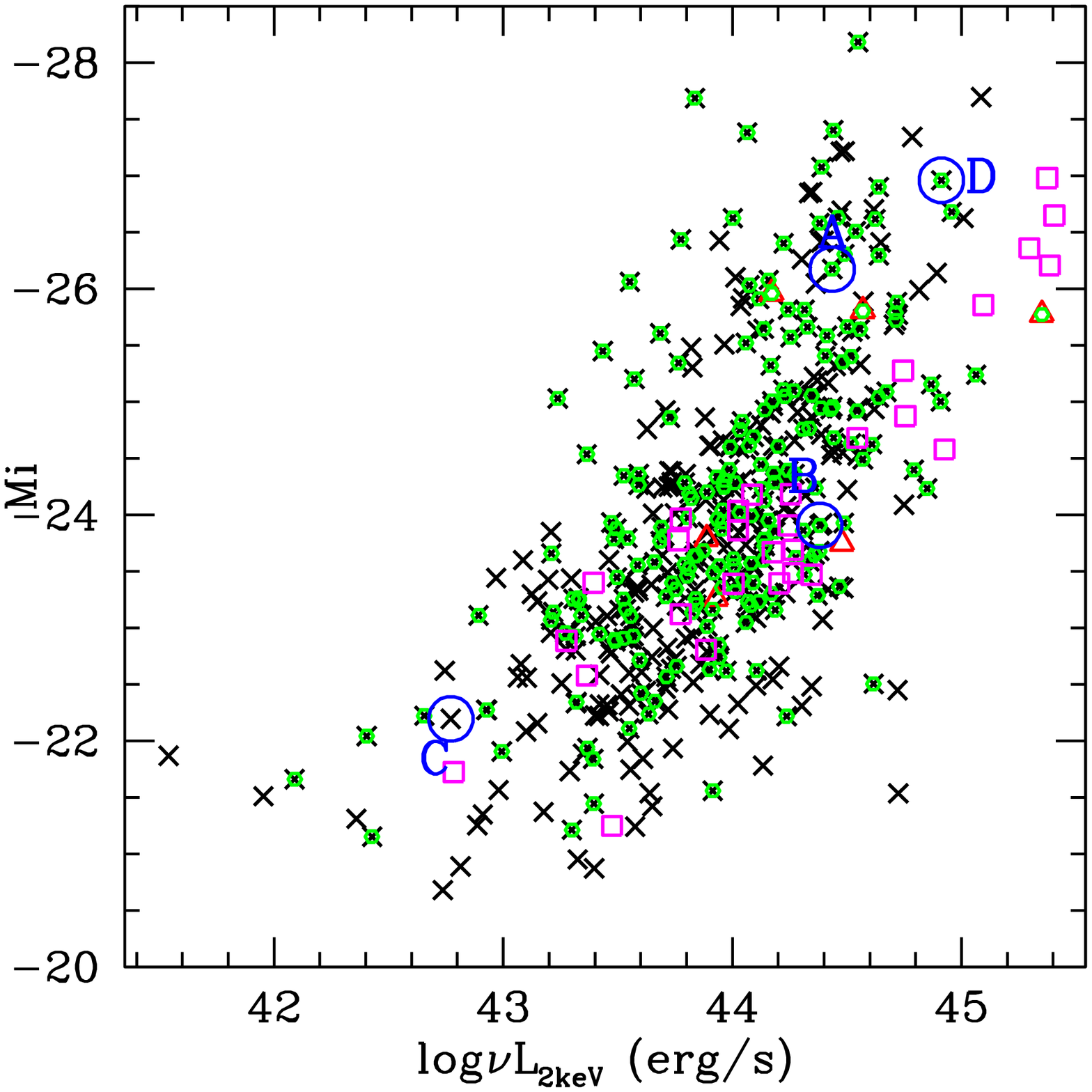, angle=0,width=\linewidth} \caption{I band
absolute magnitude versus X-ray luminosity for the 413 type 1 AGN in
XMM-COSMOS. The symbols are as in Figure~\ref{Lxz}.\label{mlx}}
\end{figure}

In Figure~\ref{Lxz}, \ref{mz} and \ref{mlx} the properties of the
XC413 sample (black crosses) are compared to the E94 quasars
(magenta squares).

Figure~\ref{Lxz} shows the rest frame 2 keV X-ray luminosity versus
redshift plane. For the XC413 quasars (black crosses), the 2~keV
luminosity is calculated from the observed X-ray band flux and the
spectral index $\Gamma$ (Mainieri et al. 2007). For the XC413
quasars without $\Gamma$ estimation, spectral index $\Gamma=2$ for
the 0.5--2keV band and $\Gamma=1.7$ for the 2--10keV band is assumed
(Cappelluti et al. 2009). The magenta squares represent the 47 E94
quasars in the same plot. The 1~keV monochromatic luminosity and the
spectral index $\alpha_x$ for each E94 quasar respectively (E94
Table 2) has been used to calculate the 2~keV monochromatic
luminosity of the E94 sample. Notice that E94 uses
$H_0=50$km~s$^{-1}$Mpc$^{-1}$, so we computed the X-ray luminosity
with the same cosmology used in this paper.

In Figure~\ref{mz}, we show the i band absolute magnitude $M_i$ --
redshift plane.For E94 we derived i band magnitudes using the V band
magnitude and (B-V), (R-I) colors from E94 (Table 4A), and the
transformation from the Johnson system to the AB magnitude in i band
given by Jester et al. (2005, Table~1; $r-i=0.90(R-I)-0.20$ and
$r=V-0.19(B-V)-0.02$). Note that we first recalculated the tabulated
E94 $M_V$ with the same cosmology used in this paper. In
Figure~\ref{mlx}, we show the location of the XC413 sample with
respect to the E94 sample in the i band absolute magnitude $M_i$
versus X-ray luminosity plane. We can clearly see that the XC413
sample covers a larger luminosity and redshift range compared to the
E94 sample.

The XC413 sample is ten times larger than the E94 quasar sample.
Over two orders of magnitude range in luminosity, the XC413 sample
homogeneously covers a redshift range (80\% quasars at $0.1\leq z
\lesssim2.2$) $\sim$6 times larger than the E94 sample (80\% quasars
at $z<0.35$), and a flux range some 2~dex fainter than the E94
sample ($16.9 \leq i_{AB} \leq 24.8$). While the two samples span
comparable range of $M_i$, the COSMOS sample includes many more
lower luminosity (log$\nu L_{2keV}<10^{43.5}~erg/s$) X-ray sources
compared to E94 sample (see Figure~\ref{Lxz}).

\subsection{X-ray Properties}

\begin{figure}
\epsfig{file=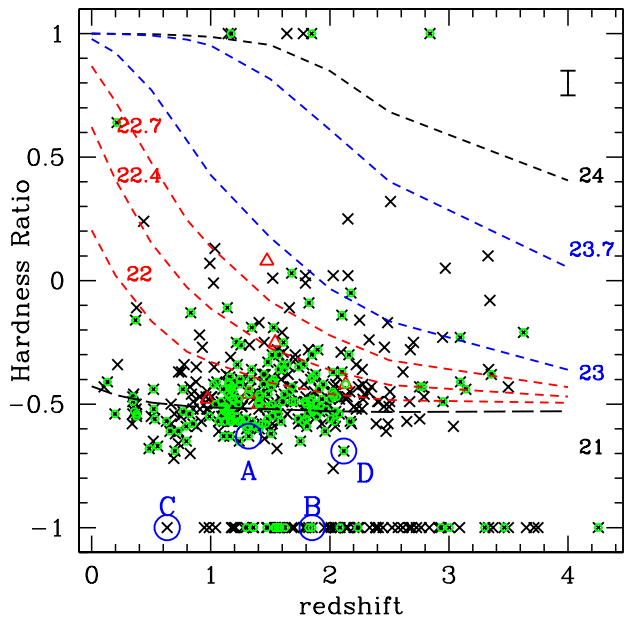, angle=0,width=\linewidth} \caption{Hardness
ratio ((H-S)/(H+S)) versus redshift for the XC413 sample. The
symbols are the same as in Figure~\ref{Lxz}. The lines show the
tracks for different column density labeled in $log(N_H)$. The line
in the upper right corner showed the size of the mean error bar of
HR. \label{hrz}}
\end{figure}

The hardness ratios of the XC413 AGN are shown versus redshift in
Figure~\ref{hrz}.  The hardness ratio is defined as HR=(H-S)/(H+S),
where H are the counts in the 2-10 keV band and S those in the 0.5-2
keV soft energy band (Brusa et al. 2010). Usually, a negative HR
indicates the absence of X-ray absorption, although for a given
spectrum the HR is a strong function of redshift. There are 73
(17.6\%) quasars with HR=-1, which is reasonable fraction for type 1
AGNs (e.g. Rosati et al. 2002). There are also 6 quasars with HR=1
in XC413, which is somewhat unexpected. We will discuss the
properties of these 6 quasars in a following paper. Figure \ref{hrz}
also shows the curves of hardness ratio versus redshift for a range
of equivalent hydrogen column densities ($N_H$), computed using
XSPEC (version 11, Arnaud 1996), assuming a spectral slope of
$\Gamma=1.7$ and solar abundances. Most(299) of the XC413 AGN have
HR values indicative of small absorbing hydrogen column densities,
N$_H <10^{22}$cm$^{-2}$, as is typical of most type~1 AGN (e.g.,
Mainieri et al. 2007). However, 27.6\% (114) of the XC413 sample
shows HR indicative of obscuration N$_H
>10^{22}$cm$^{-2}$ and even larger. Mainieri et al. (2007) estimated
the N$_H$ value for 378 quasars of the XC413 from the X-ray
spectrum. Only 36 out of the 378 quasars have N$_H$ value larger
than $10^{22}$cm$^{-2}$. For the 114 quasars lie above the N$_H
=10^{22}$cm$^{-2}$ curve in the HR versus redshift plane, 19 quasars
have estimated N$_H>10^{22}$cm$^{-2}$, 76 quasars have estimated
N$_H<10^{22}$cm$^{-2}$ and 19 quasars without the N$_H$ estimation
from the X-ray spectrum. From the Figure~\ref{hrz}, we can also see
that majority of the sources above the N$_H =10^{22}$cm$^{-2}$ curve
cluster near the curve within the range of the typical HR error bar.
A similar results have been found by Lanzuisi et al. (2012) for the
bright sources in the C-COSMOS survey (Elvis et al. 2009).

A deeper analysis of the X-ray properties of this sample, including
the optical to X-ray slope analysis, has been reported in Lusso et
al. (2010) and Brusa et al. (2010).

\subsection{Radio Loudness}
\label{s:radio}

Quasars are often classified into radio-loud and radio-quiet, based
on their radio properties. Typically, 10\% of all quasars are
radio-loud (e.g. Kellermann et al. 1989; Urry \& Padovani 1995;
Ivezi\'{c} et al. 2002).

There are several ways to classify quasars as radio-loud or
radio-quiet. We tried: (1) $R_L = log(f_{5GHz}/f_{B})$ in the rest
frame (Wilkes \& Elvis 1987, Kellermann et al. 1989), for which
$R_L>$1 defines a radio loud source; (2) $q_{24}$ = $log(f_{24\mu
m}/f_{1.4GHz})$ in the rest frame and observed frame (Appleton et
al. 2004), for which $q_{24}<0$ is defined as radio-loud; (3)
$R_{1.4} =log(f_{1.4GHz}/f_{K})$ in the observed frame, for which
$R_{1.4}>$1 defines a radio loud source; (4)
$R_{uv}=log(f_{5GHz}/f_{2500\AA})$ in the rest frame (e.g., Stocke
et al. 1992; Jiang et al. 2007), for which $R_{uv}>$1 defines a
radio loud source; (5) $log_{10}[P_{5 GHz}(W/Hz/Sr)]>24$ in the rest
frame (Goldschmidt et al. 1999) as radio-loud; (6) $R_X=\log(\nu
L_{\nu} (5GHz)/L_X)>-3$ in the rest frame (Terashima \& Wilson 2003)
as radio loud.

After cross-checking with all the above criteria, there are 6
sources classified as radio-loud in all cases\footnotemark. We
define these 6 of the 413 sources as radio-loud sources.
\footnotetext{XID=40, 2282, 5230, 5275, 5517, 5395.}
For the most commonly used criteria, using the $R_L$ criterion,
there are 17 radio-loud quasars. Using the $q_{24}$ criterion, there
are 13 radio-loud quasars. Using both $R_L$ and $q_{24}$ criteria,
gives 8 radio-loud sources\footnotemark.
\footnotetext{The two additional sources using $R_L$ and $q_{24}$
criteria are XID=5497, 54541.}
The fraction of radio-loud sources in the XC413 sample is no more
than 4.5\% using any one criterion, and no more than 2.1\% using two
or more criteria, smaller than the $\sim$10\% seen in typical
optically selected AGN samples (e.g. Peterson et al. 1997). The VLA
detection is deep enough that for great majority of the quasar in
XC413 (95.7\%) the VLA data would detect a radio-loud quasar, using
any standard criteria. The radio-loudness of the XC413 sample and
the effects of the detection limits in radio to this fraction will
be discussed in a later paper (Hao et al. 2012c, Paper IV).

\section{Spectral Energy Distributions}

Our goal is to produce uniform rest frame SEDs for all the
XC413 quasars. To do so, several corrections to the observed SEDs
need to be considered. The corrected SEDs can then be converted to a
uniform grid in the rest frame to calculate the mean and dispersion.

\subsection{Corrections to the Observed SEDs}

Several complicating factors need to be considered before we can
study the SED of the individual sources: (1) Galactic extinction,
(2) variability, (3) emission line flux contamination; (4) host
galaxy contamination. In this subsection, we will discuss the first
three effects, deferring the host galaxy correction, which is more
complex and requires more assumptions, to \S \ref{s:host}.

\subsubsection{Galactic Extinction}
\label{s:ebvcor}

\begin{figure}
\epsfig{file=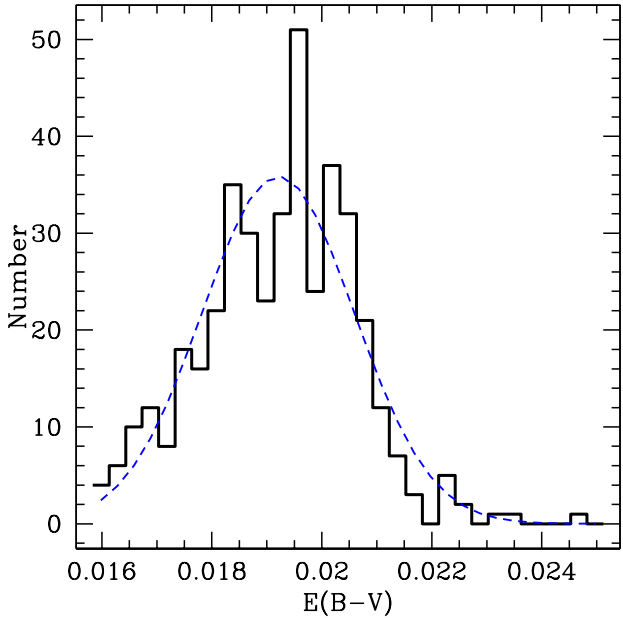, angle=0, width=\linewidth} \caption{Distribution
of estimated Galactic extinction for the XC413 AGN. Blue dashed line
= Gaussian fit. \label{ebvhist}}
\end{figure}

Although the Galactic extinction is small in the COSMOS field, with
a median of E(B-V) = 0.0195 (Capak et al. 2007), we include the
correction on a source-by-source basis to eliminate this factor. A
photometric correction for each band of the source is calculated
from the Galactic extinction of this source multiplied by the
filter-dependent factor $k$ given in Table 1.  The factors are
calculated by integrating the filter response function against the
Galactic extinction curve from Cardelli et al. (1989).

The estimated E(B-V) from Schlegel et al (1998) for the XC413 AGN
has a mean of 0.0192 and a standard deviation of 0.0014. For the
most affected band -- the FUV band, the mean Galactic extinction is
$\sim0.17\pm0.01$ mag, which reduces to $0.10\pm0.03$ mag for the
shortest wavelength optical band (the CFHT u-band). A histogram of
E(B-V) for the sample is shown in Figure \ref{ebvhist}.

\subsubsection{Variability}
\begin{figure}
\epsfig{file=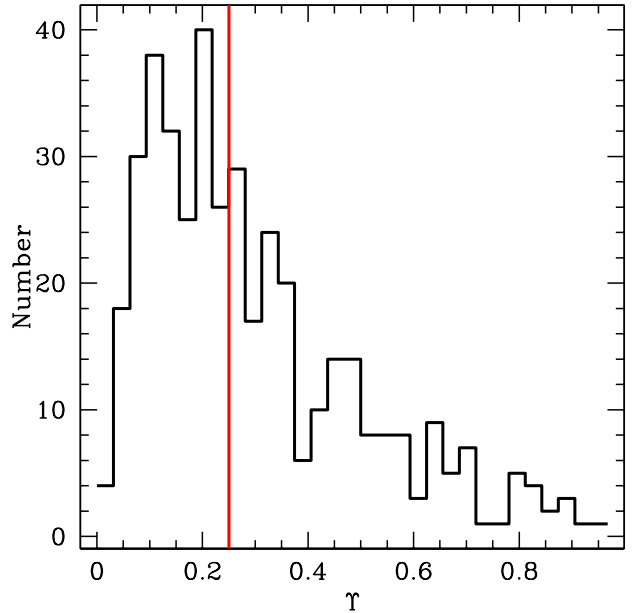, angle=0, width=\linewidth} \caption{Distribution
of the variability parameter $\Upsilon$ (Salvato et al. 2009) of the
XC413 AGN.  The vertical line divides the sources into variable
($>$0.25) and not-variable ($<$0.25). \label{varhist}}
\end{figure}

\begin{figure*}
\includegraphics[angle=0,width=0.5\linewidth]{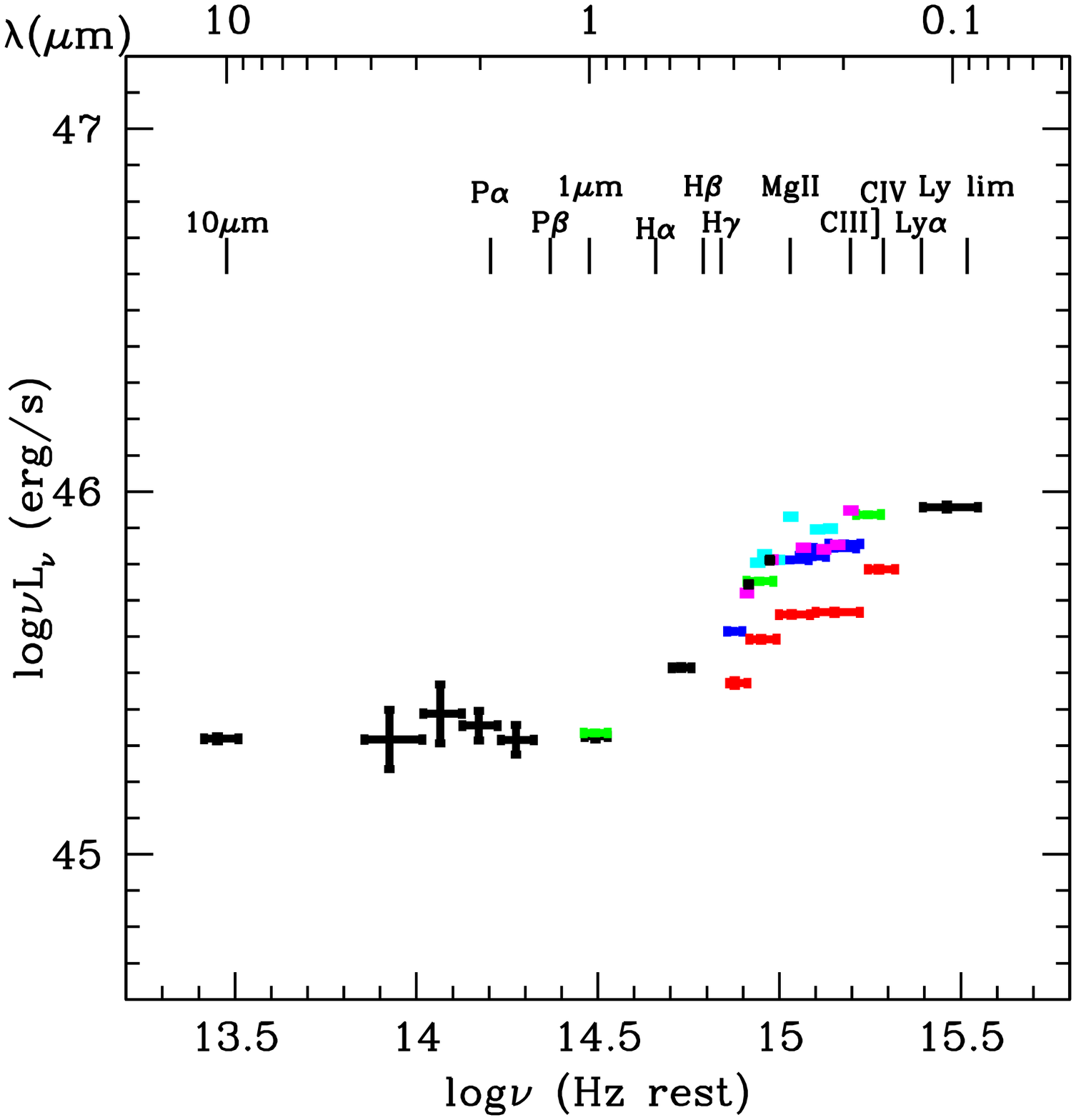}
\includegraphics[angle=0,width=0.5\linewidth]{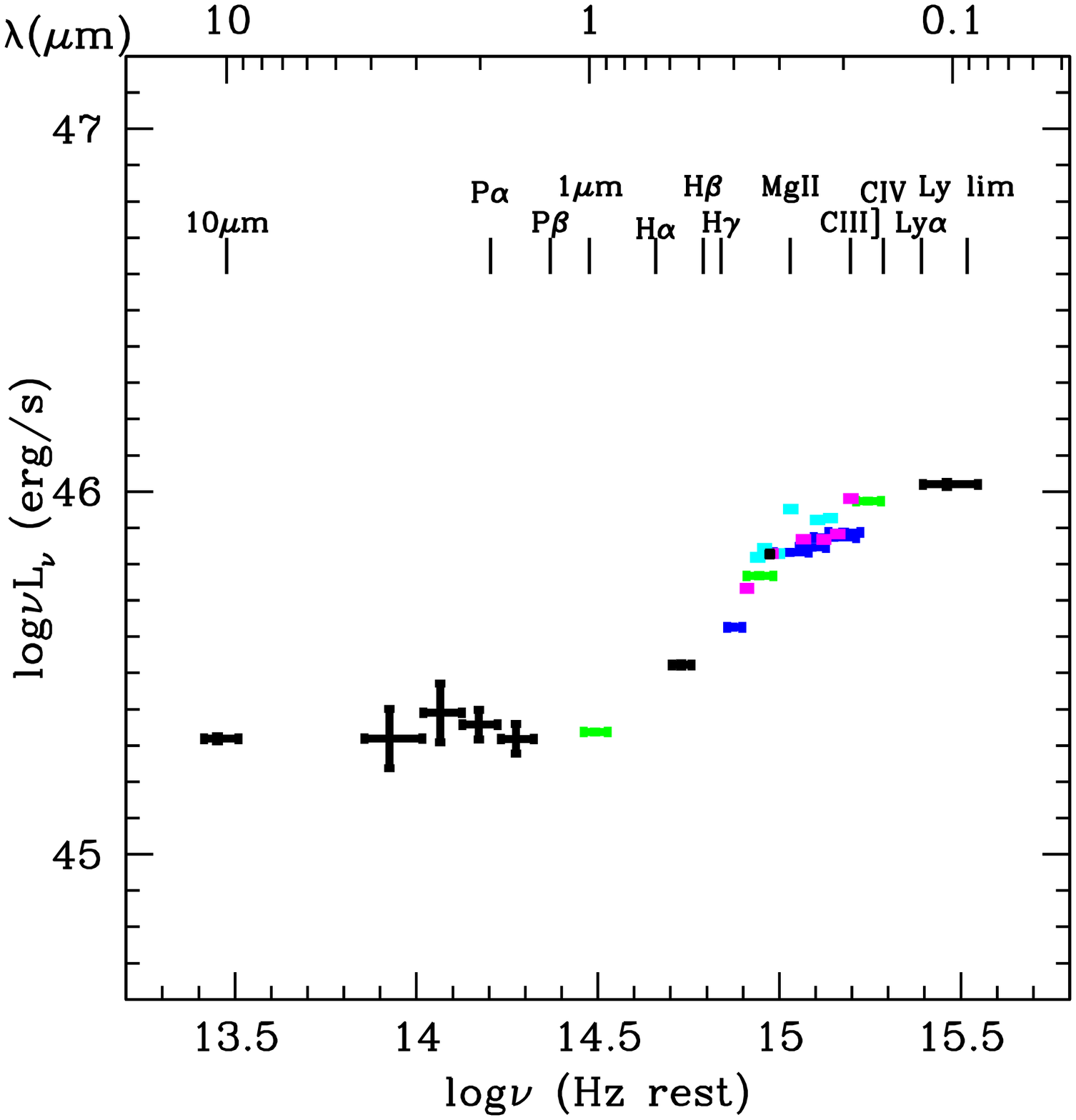}
\caption{Type 1 AGN SED showing clear flux and slope variability in
the optical/UV. The source is COSMOS\_J149.85194+1.99845 (XID=17) at
z=1.236, (i-band absolute magnitude -26.3). {\em Left:} SED
resulting from using all the data (2001-2007).  {\em Right:} only
the data in the years 2004-2007. The data points are: black, from
low frequency to high frequency, are: 24$\mu m$, 8$\mu m$, 5.7$\mu
m$, 4.5$\mu m$ and 3.6 $\mu m$, K-band, H-band, J-band and the FUV
and NUV. The red points show the broad SDSS ugriz bands from 2001.
The blue points are the Subaru broad bands from 2005. The green
points are the (CFHT) K-band, and the (CFHT) u band and i band. The
purple points are the 6 Subaru intermediate bands for season 1
(2006) (IA427, IA464, IA505, IA574, IA709, IA827).  The cyan points
are the 5 Subaru intermediate bands for season 2 (2007) (IA484,
IA527, IA624, IA679, IA738, IA767). \label{vareg}}
\end{figure*}

\begin{figure}
\epsfig{file=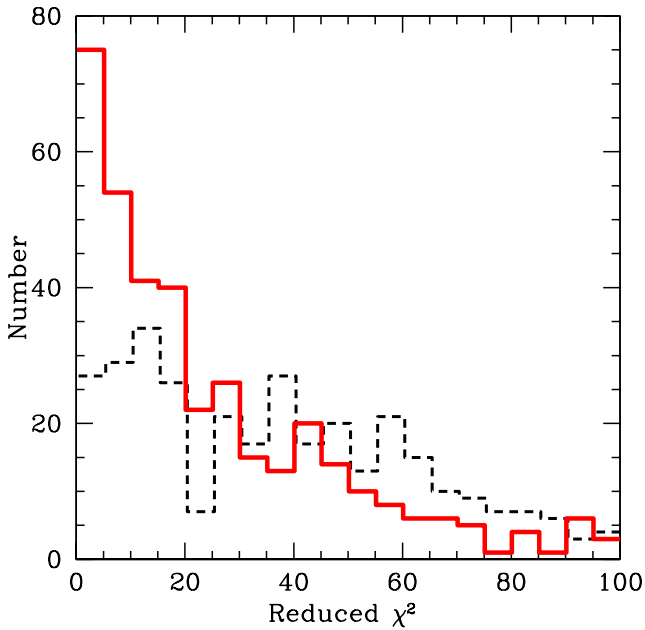, angle=0, width=\linewidth} \caption{The reduced
$\chi^2$ of the SED-fitting (see \S \ref{s:fit}) both before (black)
and after (red) the restriction of the data to the 2004-2007
interval. \label{rchisq}}
\end{figure}

\begin{figure}
\epsfig{file=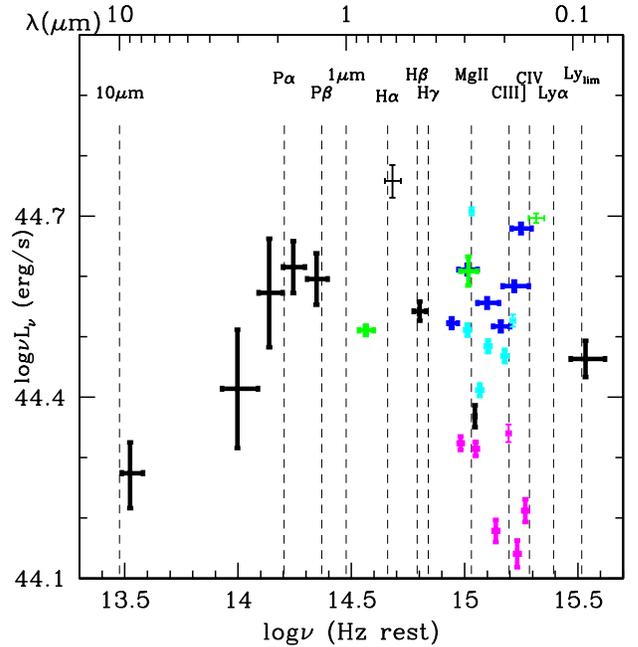, angle=0, width=\linewidth} \caption{ SED
(restricted 2004-2007 interval) of source COSMOS\_J150.48437+2.16204
(XID=2232) at z=1.641 (i-band absolute magnitude -23.5). The
inconsistent photometry, notably in the UV, suggests more rapid
variability in this objects, and in the 5\% of similar sources, than
is typical of the majority of the population. The points are color
coded as in Figure~\ref{vareg}. \label{bad}}
\end{figure}

The COSMOS optical and infrared data were taken over a 4 year
interval, from 2004 to 2007, and the SDSS data for the field were
taken as early as 2001. This long time span causes problems in
deriving the true shape of the SEDs, because AGN are significantly
variable on these timescales (Hawkins 2007, Sergeev et al. 2006).
Most AGN vary in their optical continuum flux on the order of 10\%
on timescales of months to years (Vanden Berk et al. 2004).

Variability is common in the XC413 sample too. Salvato et al. (2009)
analyzed the variability of all the XMM-COSMOS X-ray sample and
defined a convenient variability parameter $\Upsilon$ (the rms of
the magnitude offsets at the sampled epochs) to quantify the
variability of the sources. Salvato et al. (2009) found that
$\Upsilon > 0.25$ efficiently separates out variable XMM-COSMOS
sources (including both point-like and extended sources). Half of
the XC413 AGN show significant variability by this criterion
(Figure~\ref{varhist}).

An example of an AGN SED from this sample (XID=17), that clearly
changes both in flux and in optical/UV slope, is presented in
Figure~\ref{vareg} (left). The resulting SED is confused when the
entire time period is included. This is a widespread problem. Figure
\ref{rchisq} shows the broad $\chi^2$ distribution obtained in
fitting a quadratic function to the SEDs of the full data set.
Details of the SED fitting used here will be discussed in \S
\ref{s:fit}.

We do not use the Salvato et al. (2009) method to correct the SED,
although it works well for deriving photometric redshifts. This is
because the method introduces an assumed smooth average SED shape to
estimate the correction to be made to each photometry band for
variability. This might bias estimates of the intrinsic SED shape.

The alternative approach which we adopt is simply to restrict the
data set to a shorter time period.  For this approach, we
experimented with several different temporal cuts, trying to find
the best compromise between contemporaneity and completeness of
coverage. Using $\chi^2$ fits to the continuum, we find that by
using only the data in the interval from 2004 to 2007 reduces the
variability issue. Beyond that, we exclude the K band from KPNO
because it duplicates the K band from CFHT, which is deeper and has
better photometry. We also exclude the data in the filter NB816
because the photometry in this band is the sum of runs distant in
time and is thus not reliable due to the variability of the quasars.

Figure~\ref{vareg} (bottom) shows how using only the 2004-2007 data
removes the scatter in the same AGN (XID=17) shown in
Figure~\ref{vareg} (top). Similar improvements are common. The solid
line in Figure~\ref{rchisq} shows how the reduced time span greatly
improves $\chi^2$ (with a distinct peak at low $\chi^2$) after
applying these restrictions. The peak reduced $\chi^2$ is still
$\sim$5, which is mainly due to two reasons: 1) the photometry of
the data is so good that deviations from the simple assumed
continuum, due e.g. to weak emission lines and blends, as well as
weak residual variability, remain significant; 2) when calculating
the $\chi^2$, we use the formal photometric errors from various
catalogs, which are usually somewhat underestimated.

The SEDs of $\sim$5\% of the sources remain scattered, as in the
example in Figure~\ref{bad}. While some of the points are strongly
affected by emission lines (see next section), even the unaffected
points seem to show $\sim$0.3~dex changes. This indicates more rapid
variability than is seen in the majority of the sample.

The 8 photometric bands excluded are starred in Table~1. We have 35
photometry bands remaining with high photometry quality. This
temporal cut gives a sample free of strong variability without
having to make any assumptions for the SED shape, and retaining
virtually the same photometric coverage.

\subsubsection{Broad Emission Line Fluxes}
\label{s:bel}

\begin{figure*}
\includegraphics[angle=0,width=0.5\linewidth]{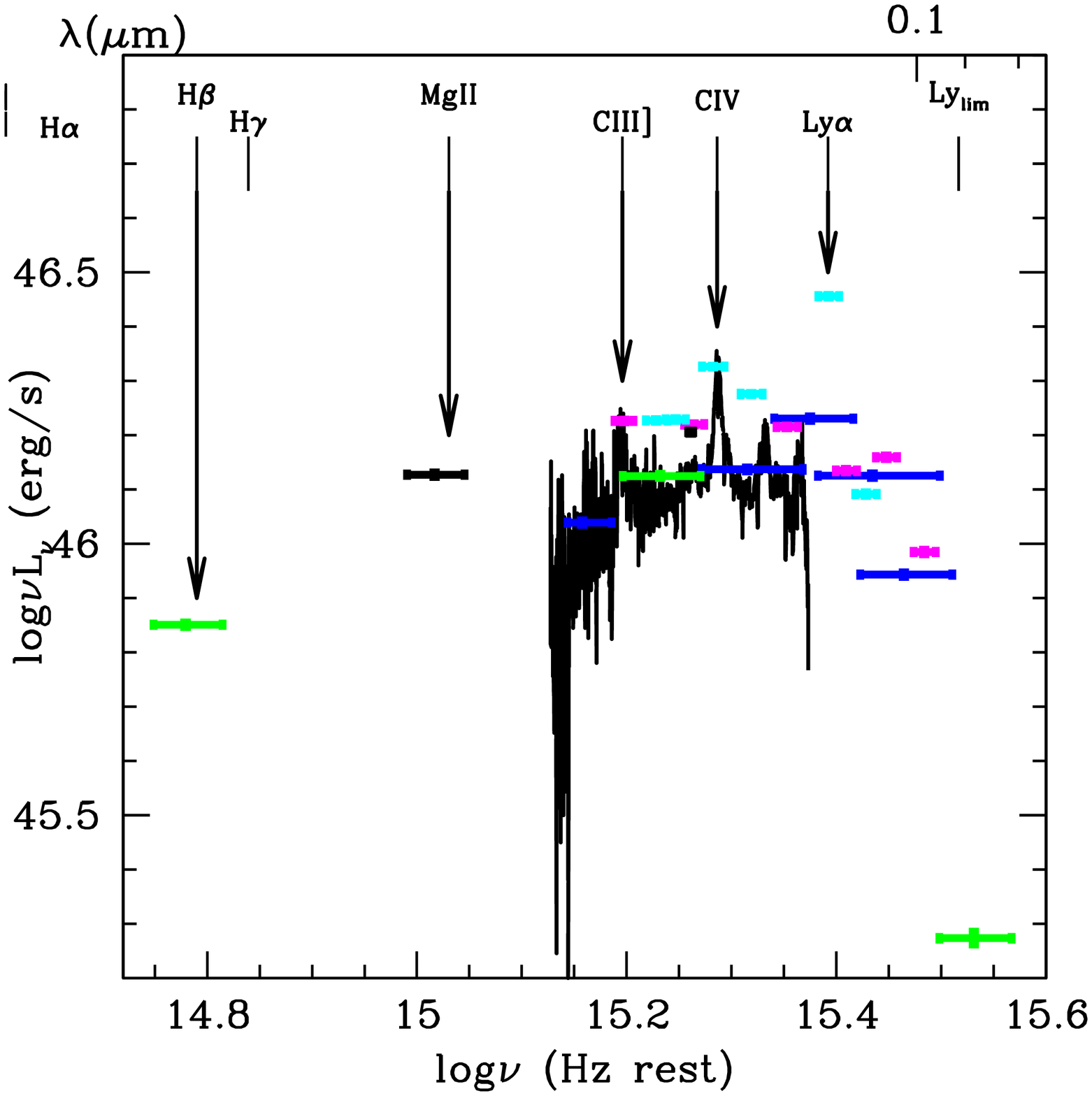}
\includegraphics[angle=0,width=0.5\linewidth]{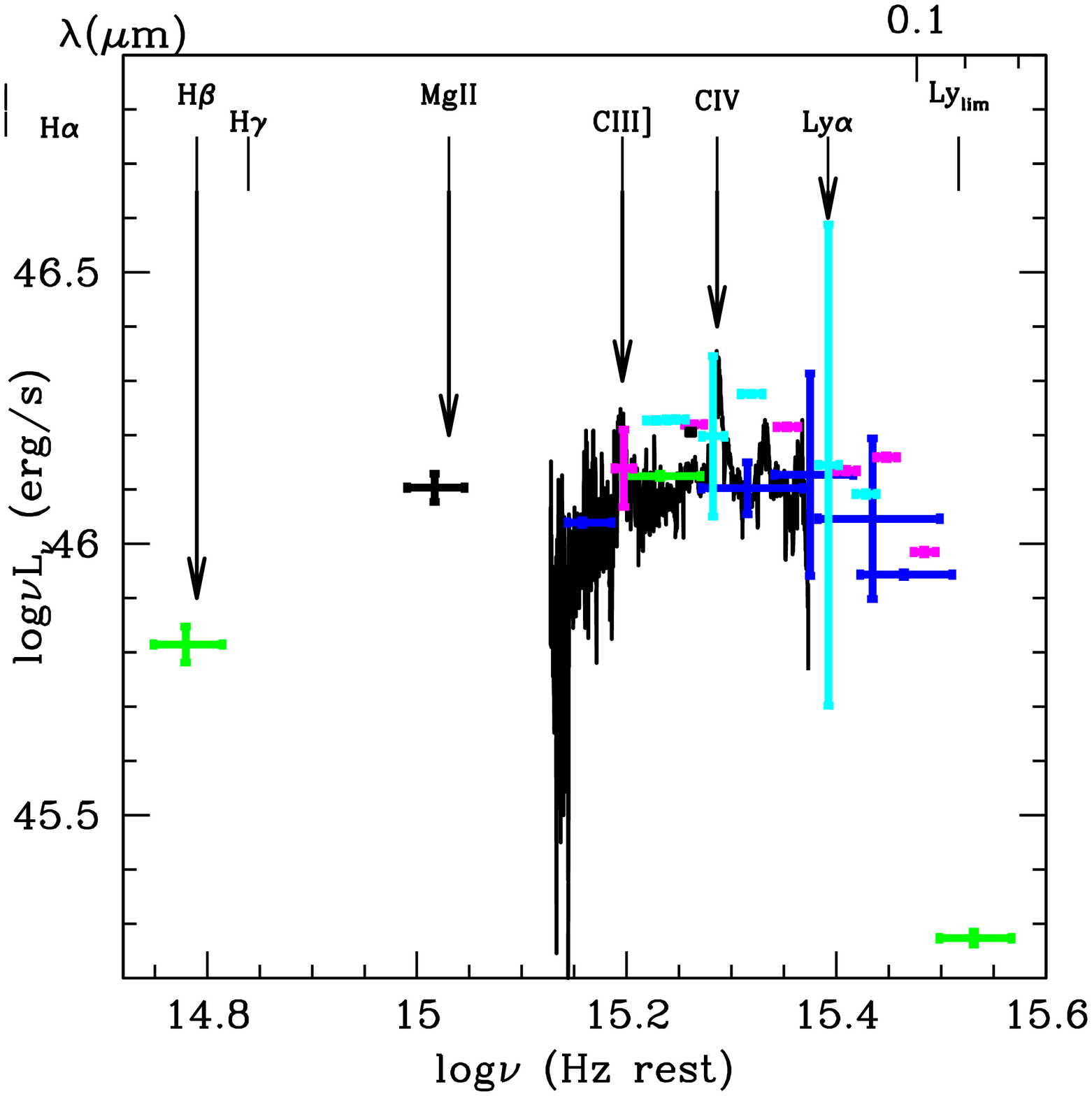}
\caption{{\em Left:} Optical SED of COSMOS\_J150.20885+2.481935
(XID=180) at z=3.333 (i band absolute magnitude -27.19), strongly
affected by BELs. Highly significant departures due to Ly$\alpha$
and \ion{C}{4} distort fits to the continuum SED by $\sim$0.3 dex
and $\sim$0.1 dex, and possible BELs of \ion{C}{3}], \ion{Mg}{2} and
H$\beta$ are also apparent. The black solid line show the VLT
spectrum of this source.
{\em Right:} Optical SED of COSMOS\_J150.20885+2.481935 (XID=180)
after removing the emission lines using the prescription in
\S\ref{s:bel}. The error bars on the affected photometry points
become larger because of the wide spread in BEL EWs
(Figure~\ref{ewhistn}). The black solid line show the VLT spectrum
of this source. The points are color coded as in Figure~\ref{vareg}.
\label{eg1}}
\end{figure*}

\begin{deluxetable*}{ccccccc}
\tabletypesize{\scriptsize} \tablecaption{Spectral Line Rest-Frame
Equivalent Width from SDSS DR7 \label{t:ew}} \tablehead{
\colhead{Spectral }& \colhead{N\tablenotemark{1}} &\colhead{Mean
\tablenotemark{2}}& \colhead{Mean \tablenotemark{3}}& \colhead
{$\sigma$ \tablenotemark{3}} &
\colhead{correction\tablenotemark{4}} & \colhead{correction\tablenotemark{5}}\\
\colhead{Line }& & \colhead{(Gaussian)} & \colhead{(Log Normal)} &
\colhead{(Log Normal)}
& \colhead {$r$} & \colhead {$IA624$}\\
&  & \colhead{EW(\AA)} & \colhead{EW(\AA)}  & \colhead{EW(\AA)}  &
\colhead{(dex)} &\colhead{(dex)}} \startdata
$Ly\alpha$        & 4   & 55.3 & 58.6 & 50.5 &  0.052 & 0.200\\
\ion{C}{4}        & 39     & 26.3 & 26.7 & 15.2 & 0.024  & 0.103\\
\ion{C}{3}]       & 70      & 17.6 & 17.6 & 6.7 & 0.016  & 0.070\\
\ion{Mg}{2}        & 137   & 20.8 & 20.7 & 8.4 & 0.019  & 0.082\\
$H\gamma$        & 21 & 13.4 & 13.5 & 6.1 & 0.013  & 0.055\\
$H\beta$            & 25   & 29.3 & 29.8 & 17.4 & 0.027  & 0.113\\
$[$\ion{O}{3}$]$4960  & 19 & 13.7 & 12.4 & 10.4 & 0.012  & 0.051\\
$[$\ion{O}{3}$]$5008  & 19 & 25.0 & 22.3 & 22.5 & 0.021  & 0.087\\
$H\alpha$           & 13 & 204.3 & 186.7 & 226.0 & 0.148 & 0.457\\
\enddata
\tablenotetext{1}{This column is the number of detections in our
optical (Magellan/MMT/SDSS) spectra for each spectral
line.}\tablenotetext{2}{Mean EW correspond to the Gaussian fit of
the histogram in Figure~\ref{ewhistn}.} \tablenotetext{3}{The mean
and sigma of the log-normal fit of the histogram in
Figure~\ref{ewhistn}.} \tablenotetext{4}{Typical correction for a
broad band for quasars at redshift $z=2$, here we use r band
(BW=1382\AA) as an example. The correction is calculated as
$log(BW/(BW+EW_{rest}*(1+z)))$.}\tablenotetext{5}{Typical correction
for an intermediate band for quasars at redshift $z=2$, here we use
IA624 (BW=300\AA) as an example. The correction is calculated as
$log(BW/(BW+EW_{rest}*(1+z)))$.}
\end{deluxetable*}

\begin{figure*}
\begin{center}
\includegraphics[angle=0,scale=0.9]{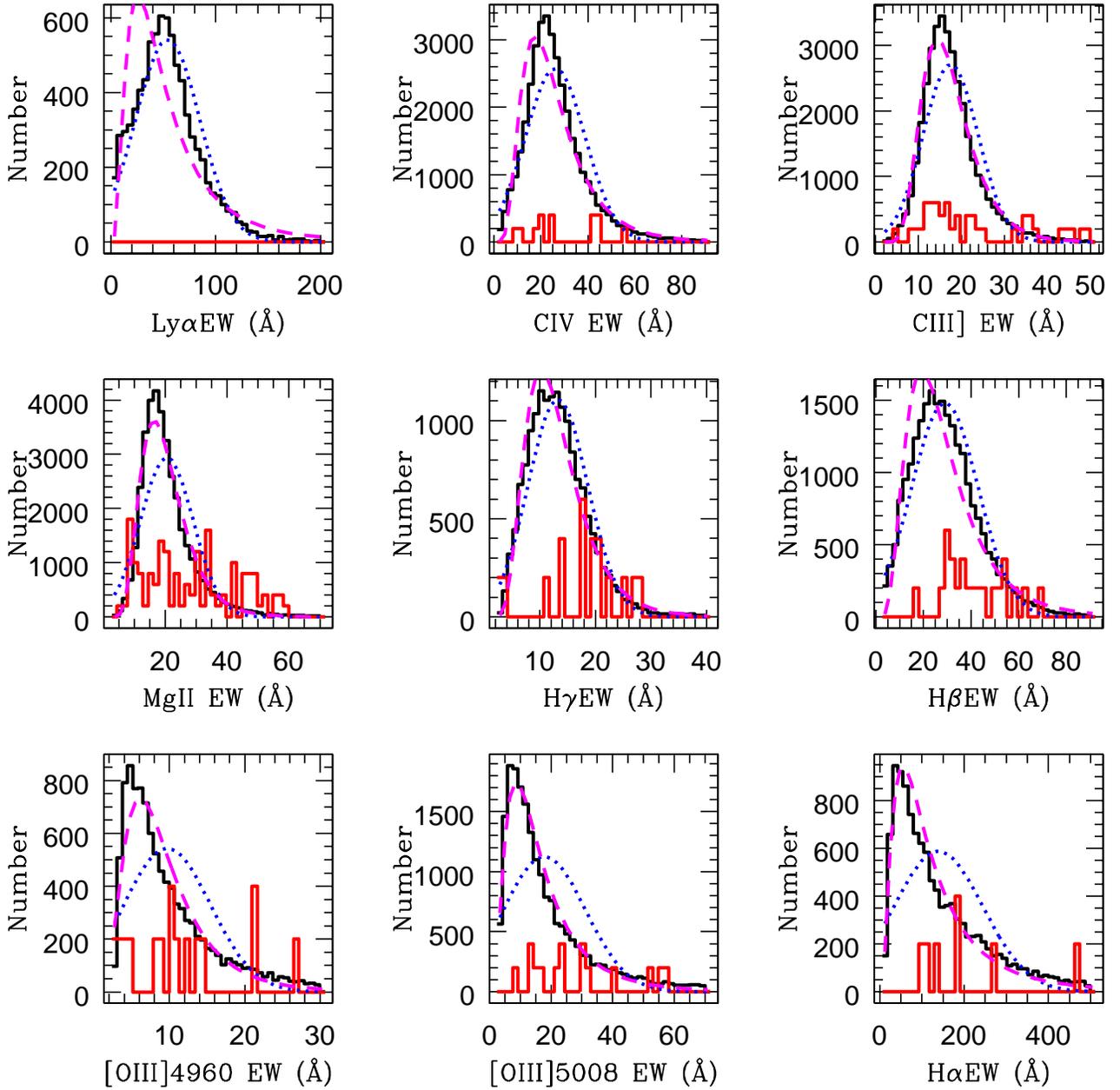}
\caption{ EW distributions of 7 BELs and the [OIII] doublet from
SDSS DR7 (black histogram). Blue dotted line = Gaussian fit; magenta
dashed line = log-normal fit. The EWs measured for the COSMOS Type~1
AGN sample is shown as a red solid histogram, multiplied by 200 for
plotting purposes. For Ly$_\alpha$, the 4 measured EWs are all
larger than the range of the histogram. \label{ewhistn}}
\end{center}
\end{figure*}

The broad emission lines (BELs) of quasars clearly affect the
photometry in numerous objects.  Figure~\ref{eg1} shows an example,
in which the broad Ly$\alpha$ emission line has increased the
intermediate band by $\sim0.3$~dex compared to the adjacent broad
band. Similarly, the broad \ion{C}{4} emission line has increased
the flux by approximately 0.1~dex. Visual inspection shows that 280
of the 413 AGN have photometry that appears clearly affected by
BELs.

To correct the SEDs for BEL contamination, we need to measure or
estimate equivalent widths (EWs) for all the strong BELs in the
quasar spectra. Although, by construction, we have optical spectra
for each of the type~1 AGN in the sample, these spectra do not span
the full spectral range required. Typically only one or two BELs are
present in each optical spectrum. Estimates of EW for most of the
BELs are still needed.

For the BELs in the spectra of IMACS/Magellan, VIMOS/VLT or SDSS
(Trump et al. 2007, 2009a; Lilly et al. 2007, 2009; Schneider et al.
2007), the EWs were measured individually using the {\tt splot}
package in IRAF\footnotemark. The number of measured EWs for the
major BELs are shown in the second column of Table~\ref{t:ew}. The
MgII BEL has measured EWs for 33\% of the sample (137/413), but no
other BEL has more than 17\% coverage.
\footnotetext{IRAF is distributed by the National Optical Astronomy
Observatory, which is operated by the Association of Universities
for Research in Astronomy, Inc., under cooperative agreement with
the National Science Foundation.}

For the remaining BELs, we must estimate their EWs by other means.
We considered three methods:
\begin{enumerate}
\item Use correlations of EW between the different BELs to bootstrap from the
  observed lines to the rest;
\item Use the known relationships between CIV EW and luminosity or redshift,
  i.e.  the Baldwin effect (Baldwin 1977), or analogs for other lines (Green et
  al. 2001).
\item Take a mean or median EW for each BEL from a survey covering a wide
  redshift range, in order to cover all the major BELs, and use the dispersion
  of the observed EW as an error bar on the estimate.
\end{enumerate}

(1) {\em BEL correlations.} This method leads to large uncertainties
on the BEL EW estimates. We studied the correlation between the EWs
of the main BELs using the complete, UV-excess selected,
Palomar-Green X-ray sample with redshift $z<0.4$ (Shang et al.
2007).  We found that some line pairs had well-correlated EWs: MgII
and CIV (correlation coefficient, $R=0.93$), Ly$\alpha$ and CIV
($R=0.86$), CIII] and CIV ($R=0.90$), CIII] and MgII ($R=0.82$),
H$\alpha$ and H$\beta$ ($R=0.91$). However, there was either no or
weak correlation between some other key line pairs: the Balmer lines
with MgII, CIV, CIII], or Ly$\alpha$, and Ly$\alpha$ with MgII
($R<0.5$). This is quite a surprising result, and may indicate a
wider variety of ionizing continua and/or BLR conditions and
geometry than has been typically considered (Ferland and Osterbrock
2005). In order to eliminate the effect of different selection
methods, we also studied the correlation between BELs for an X-ray
selected survey, the RIXOS survey (Puchnarewicz et al. 1997). We
found no strong correlation between {\em any} of the BELs ($R<0.4$)
in the RIXOS survey.

(2) The {\em Baldwin effect and its analogs} can be studied using
EWs from the Large Bright Quasar Survey (LBQS) sample (Green et al.
2001). Green et al. (2001, Figures 1, 4) show that the widths of the
correlations between BEL EW and either monochromatic luminosity at
2500~\AA \ or redshift is $\sim$1~dex, too large to provide useful
EW estimates.

(3) The {\em EW mean and dispersion for each BEL}, require a large
sample to give well-determined values. The obvious choice is the
SDSS spectroscopic sample of quasars.  We used the EWs of the seven
main BELs and the narrow [\ion{O}{3}] doublet (Table~2) from SDSS
DR7 (Abazajian et al. 2009), as retrieved from the SDSS
archive\footnotemark.
\footnotetext{http://www.sdss.org/dr7/products/spectra/index.html}
We do not directly use the composite quasar spectrum from SDSS
(Vanden Berk et al. 2001), because the EW distribution dispersion for each BEL
is not available.

For each quasar, we used only those lines for which the DR7
indicated a good fit ($nsigma>10$), which amounts to about 65\% of
the SDSS type 1 AGN. As the cataloged EW are in the observed frame,
we transformed them to the rest frame by dividing them by the
$(1+z)$, where $z$ is the corresponding redshift of the SDSS quasar.
Figure~\ref{ewhistn} shows the resulting distributions of the
rest-frame equivalent width of the emission lines we consider sorted in
wavelength. A Gaussian is a reasonably good approximation for
the Ly$\alpha$, and to a lesser extent, H$\beta$ and H$\gamma$
distributions. However, CIV, CIII], MgII, [OIII], and H$\alpha$ show
quite skewed distributions, with a tail to large EW, so that a
Gaussian is a poor approximation. For these lines we instead fit a
log-normal distribution to the histograms. This produced much better
fits (Figure~\ref{ewhistn}). The resulting mean and sigma of the
log-normal distribution are given in Table~\ref{t:ew}.

Comparing these three methods, we find that the mean and dispersion
of the EW from SDSS provides the best correction.

We adopted the mean and sigma from Table~\ref{t:ew} as the mean
rest-frame EW and the error bar in making the BEL subtraction from
the SEDs. We made the following simplifying assumption to calculate
the BEL correction: when the BEL central wavelength lies within the
photometry band, which is equivalent to having more than 1/2 the BEL
flux in the filter assuming symmetry, we consider them as if the
entire line profile were in the range. When the central BEL
wavelength is outside the filter band, we assume the contribution of
the broad emission line is negligible.

We need to transform the rest-frame EW back to the observed EW of
the quasar in XMM-COSMOS sample $EW_{obs}=EW_{rest}\times(1+z)$.
Here $z$ is the redshift of the quasar undergoing the correction.
The EW uncertainties were added to the photometric errors according
to the formula below.

Assuming that the continuum flux $f_c$ is constant across each broad
band photometry interval, then the integrated total flux ($\nu
f_{\nu}$) in the broad band photometry range is $F_O$. If the
bandwidth is BW, it is easy to see that $F_O=f_c\times
EW_{obs}+f_c\times BW$, based on the definition of EW. The
integrated continuum flux can then be calculated as,
 \begin{equation}
 F_c=f_c\times BW=\frac{F_O*BW}{EW_{obs}+BW}=\frac{F_O\times BW}{EW_{rest}\times(1+z)+BW}
 \label{e:fc}
 \end{equation}
so the correction that needs to be applied to the $log\nu f_{\nu}-log\nu$ diagram
is $log(BW/(EW_{obs}+BW))=log(BW/(EW_{rest}\times(1+z)+BW))$. The error on the
integrated continuum flux is
 \begin{eqnarray}
d(log\nu  f_{\nu})&=&d(log \nu f_O)+d(EW_{obs})/(EW_{obs}+BW)\nonumber\\
&=&d(log \nu f_O)+\frac{d(EW_{rest})(1+z)}{EW_{rest}\times(1+z)+BW}
 \label{e:efc}
 \end{eqnarray}

As can be seen from the above formula, the correction for the BEL
emission is, naturally, larger if the bandwidth is smaller. For
broad band photometry, the filter width is of order $\sim$1000~\AA,
while the observed EW of the BELs is of order $\sim$100~\AA\ for the
strongest lines, leading to a $\sim$10\% correction to the
measurement of the continuum, according to equation (\ref{e:fc}).
Assuming the error of the measurement of the observed EW is the
square root of the measurement ($\sim$10~\AA), the error caused by
the correction is then $\sim$1\%, according to equation
(\ref{e:efc}). The Subaru intermediate bands have filter widths of
$\sim$200~\AA\ (Table~1), so the correction for these bands is
$\sim$3 times larger.  To illustrate how large the correction is,
the corresponding corrections for an intermediate band (IA624) and
broad band (r band) are shown in Table~\ref{t:ew}. Note that no
H$\alpha$ line would lie in these two bands.

As a final step we checked that the correction was reasonable by
visual inspection. We find that 217 of the 286 AGN with clear BEL
contributions appear to be well-corrected by the scheme adopted,
while 21 appear over-corrected and 48 still appear to have residual
BEL emission in their SEDs. The error bars at the positions of the
excised BELs are large in the corrected SEDs, due to the additional
correction error, which means these affected bands will weigh less
in the SED fitting.

Note that previous SED studies (e.g., E94, Richards et al. 2006) do
not include the BEL corrections as we did here, because the SEDs in
those studies were all constructed using broad band photometry,
which is affected only at the 1\%--5\% dex level by the BEL
contribution (Table~\ref{t:ew}, column 6). For H$\alpha$, which
would only be found in the i band and beyond (typically J H or K
bands), the effect will be less than 7\% dex. Instead in COSMOS,the
majority of the optical SED comes from the Subaru intermediate bands
(season 1 and 2, 12 bands), which are affected by the BEL
contribution at 5\% to 20\% dex level (Table~\ref{t:ew} last
column). To minimize the confusion to the SED shape from the
intermediate bands by this effect, the BEL correction is thus
necessary for XC413 sample.

\subsection{Conversion to a Uniform Rest Frame Grid}

\begin{figure*}
\includegraphics[angle=0,width=0.5\linewidth]{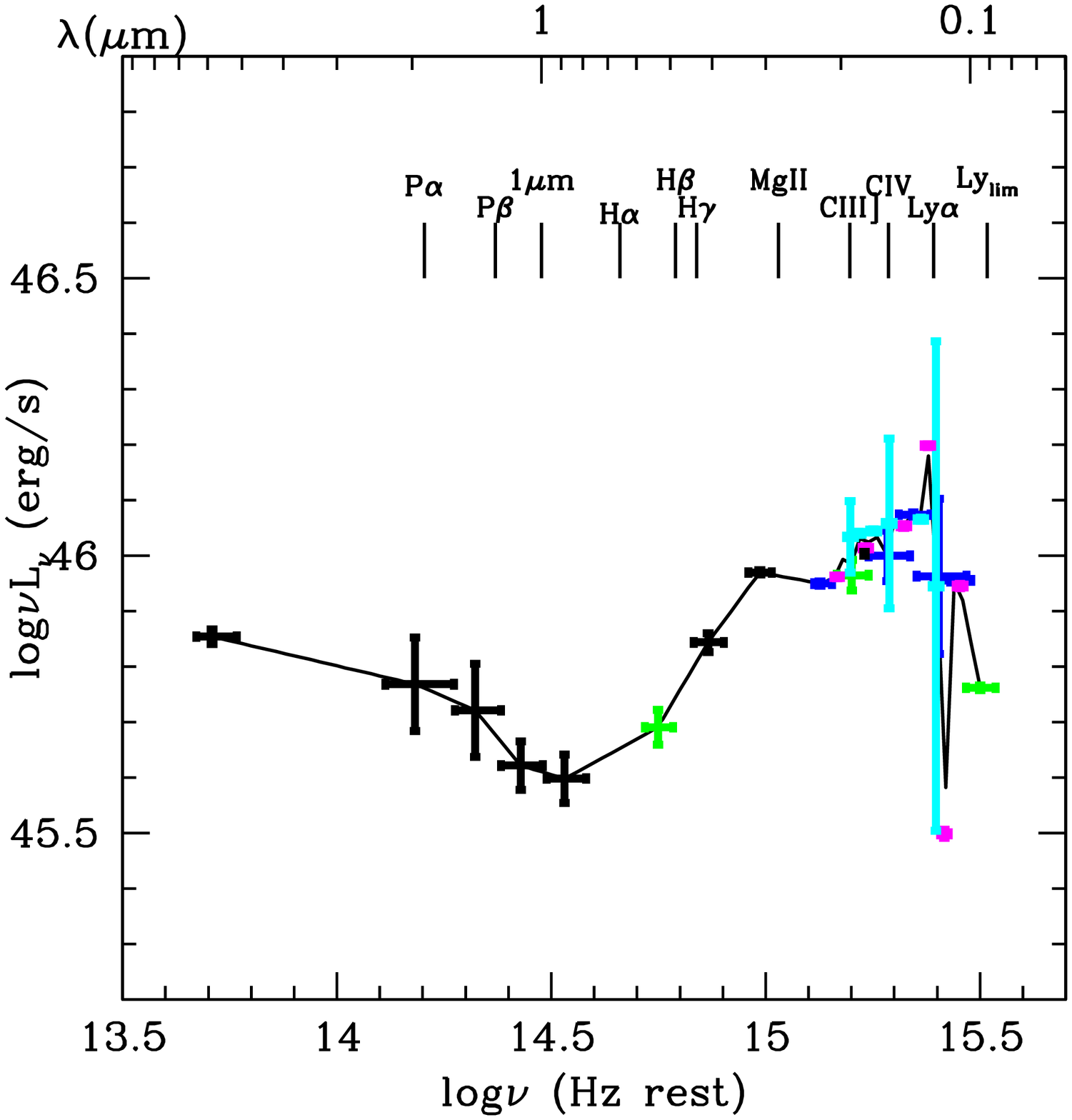}
\includegraphics[angle=0,width=0.5\linewidth]{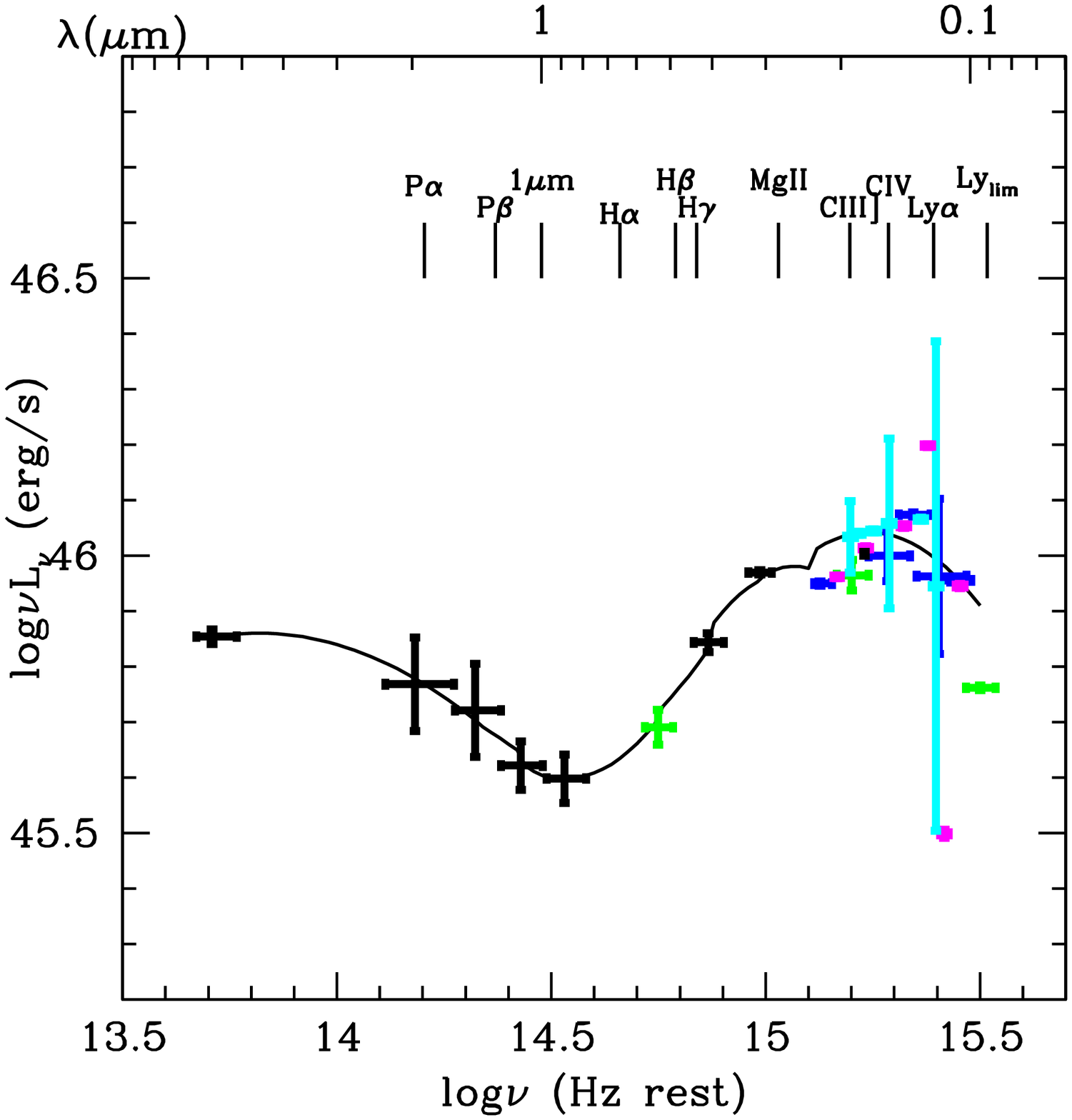}
\caption{{\em Left:} SED of COSMOS$_J$150.6088+2.76966 (XID=5331)at
z=3.038, with i band absolute magnitude -26.84, after the broad
emission line and variability (by restricting the dates of the data
set) corrections. The black line shows a linear interpolation
through the corrected data set. {\em Right:} SED of
COSMOS$_J$150.6088+2.76966 (XID~5331). The black line shows the
quadratic fit to the corrected data set. The reduced $\chi^2$ is
145.1. The points are color-coded as in Figure~\ref{vareg}.
\label{egf}}
\end{figure*}

The redshifts of the sample quasars range from 0.1 to 4.3, so any
observed photometry point spans a wide range of rest frame
frequency.  To study the individual SEDs, and to calculate a mean
SED, we clearly need to shift the SEDs of the quasars to a common
rest frame frequency grid.  For ease of comparison to E94, we
adopted the grid of the E94 mean SED, which has points separated by
0.02 in logarithmic frequency.

We first converted the flux densities at each frequency for each
object to luminosity, and shifted the observed frequencies to the
rest frame for each source. We then tried two techniques for
interpolating the observed photometry to the uniform grid points:
linear interpolation and polynomial fitting. Both methods have
advantages and disadvantages. In section \ref{s:fullsed}, we will
see that the resulting mean SED using both methods agree well with
each other with difference less than 0.02 dex in rest frame 0.1--10
$\mu m$ (Figure~\ref{msedcomp}). The wavelength/frequency discussed
in this section are all in rest frame.

\subsubsection{Linear Interpolation}
\label{s:interpolation}

The simplest way to produce a uniformly sampled SED is to linearly
interpolate between the data points in $log\nu L_{\nu}$ versus
$log\nu$ space (i.e. connecting the individual points with a power
law in linear space).

The COSMOS photometry for the XC413 sample is $>$90\% complete from
$u$ (CFHT) to MIPS 24$\mu$m (Table 1). The apparent drop in the $i$
band from Subaru is due to saturation; the remaining objects are
picked up in the $i$ (CFHT) data. The H-band (Calar Alto) is only
61\% complete, but the neighboring J, K data is complete making
interpolation straightforward. So over the 1.8 dex wide 0.35$\mu$m
-- 24$\mu$m ($\sim$0.14$mu$m -- 10$\mu$m for the typical z = 1.5 of
XC413) observed frame interpolation is unproblematic.

For the mid-infrared part of the SED the 70$\mu$m and especially the
160$\mu$m detections become sparse (8\% and 2\% respectively).  For
the detections we joined the 24$\mu$m data to the longer wavelength
points with a power-law in log~$\nu$f$_{\nu}$ vs. log~$\nu$ space.
For non-detections we extrapolated from the rest frame 24$\mu$m to
8$\mu$m slope. We checked if the extrapolation generally works by
checking the SEDs of the 34 quasars with MIPS70 detection (3 of them
also have MIPS160 detection). We found that only for 17.6\% quasars
(6 out of 34), the extrapolation of the 24$\mu m$ to 8$\mu m$ exceed
the observations. For half of the 6 quasars, the deviation is within
$1\sigma$, which means the extrapolation only fails in 8.8\% of the
case. So we use this extrapolation for all the XC413 quasars without
$70\mu m$ and 160$\mu m$ detections.

In the sub-mm/mm band from 100 GHz (3 mm) to $160\mu m$ part, the
SED can be approximated by the red end of the grey body
$f_{\nu}\propto \nu^{3+\beta}/(e^{h\nu/kT}-1)$, when $h\nu\ll kT$,
$f_{\nu}\propto \nu^{2+\beta}$ (e.g. Lapi et al.  2011). $\beta$ is
generally chosen in the range 1--2 (Dunne \& Eales 2001), and we
used $\beta=1$. I.e. we assume a power-law $f_{\nu}\propto \nu^{3}$
in this band.

For the radio band, for each source with a $>3\sigma$ detection, we
assumed a power law $f_{\nu}\propto \nu^{-0.5}$ (e.g., Ivezi\'{c}
2004) in the rest frame 1.4 GHz (21 cm) to 100 GHz (3 mm) range. The
radio power is never a significant contributor to the total
luminosity ($<$3\% Hao et al. 2012a). For sources with only a radio
upper limit, we extend the power-law of $f_{\nu}\propto \nu^{3}$ in
the sub-mm/mm to the radio wavelength.

Turning to the high frequency SED: The bluest band in the optical
with complete coverage is the u band at 0.38 $\mu m$. In the
ultraviolet the COSMOS photometric completeness drops to 60\% at
0.23$\mu m$ (GALEX NUV) and 33\% at 0.15$\mu m$ (GALEX FUV). The FUV
points would add little to the SED study as FUV fluxes are strongly
reduced by the Lyman-$\alpha$ forest induced break at 0.12$\mu$m for
z$\gtrsim$0.3, i.e. the great majority of the sample (408/413=99\%).
The Lyman-$\alpha$ forest affects the NUV photometry for z$>$0.9
(358/413=87\% of XC413). Almost all (146/150) the quasars without
NUV detection are at z$>$0.9 and all the quasars without FUV
detection are at z$\gtrsim$0.3.

For the X-ray band, we used the photon spectral index ($\Gamma$) for
the 318 objects in Mainieri et al (2007). For the remainder we
assumed flat X-ray spectrum $\Gamma$=2 (Cappelluti et al. 2009). The
observed 0.5~keV--10~keV fluxes correspond to a rest-frame band of
1.25~keV--25~keV at z = 1.5. We thus extrapolated the observed flux
to get the luminosity in the rest-frame 0.5~keV-40~keV band using
the power law slope $2-\Gamma$ in the log$\nu$f$_{\nu}$ versus
log$\nu$ space.

To join the soft X-rays to the UV through the unobservable EUV band
we directly connect the detections in the optical or uv to the
0.5~keV flux with a power-law in log$\nu$f$_{\nu}$ versus log$\nu$
space in keeping with the findings of Laor et al. (1997).

Although linear interpolation is simple, this method could be
inappropriate for several reasons:

\begin{enumerate}
\item The optical data are rich, and overlapping of bands occurs frequently in
this range. The data were taken over a 4 year time interval, so that fluxes in
adjacent and overlapping bands can be discontinuous;
\item The EW correction for the BELs is not perfect, which could drag the
  `continuum' to deviate from the correct value;
\item The fluxes in different filters have different error bars and so should be weighted
  differently, which cannot be done by interpolation.
\end{enumerate}

These factors can lead to poor continuum fits when a linear
interpolation is used, and also does not make use of the information
contained in the error bars.

\subsubsection{Polynomial Fitting}
\label{s:fit}

An alternative to the linear interpolation is to make a weighted fit
of the data using a low order polynomial. We tried several curve
fitting methods and found that a simple least squares quadratic fit
works well.

For each source, we fit quadratic functions to the observed data
from rest frame 9000~\AA\ to the Lyman limit (912\AA). The longer
wavelength data (1--160$\mu$m) were interpolated with a second
quadratic function. We smooth the junction of the two quadratics by
quadratic interpolation with a smoothing window of 10 grid points,
corresponding to a factor of 1.6 in frequency. The disadvantage of
this fitting method is that a particular functional form is assumed,
which may not be a good representation of the true SED shape of the
AGN. Just over 50\% of the sources gave reduced $\chi^2<$20, showing
that additional structure, or residual variability, is likely to be
present. An example of the linear integration and the quadratic fit
is shown in Figure \ref{egf}. For this quasar, the reduced
$\chi^2=145.1$.

The resulting SEDs for each of the objects in the XC413 are
available as VO compatible FITS files in the on-line version of the
paper.

\subsection{Full Sample Mean SED and Dispersion}
\label{s:fullsed}

\begin{figure*}
\includegraphics[angle=0,width=0.5\linewidth]{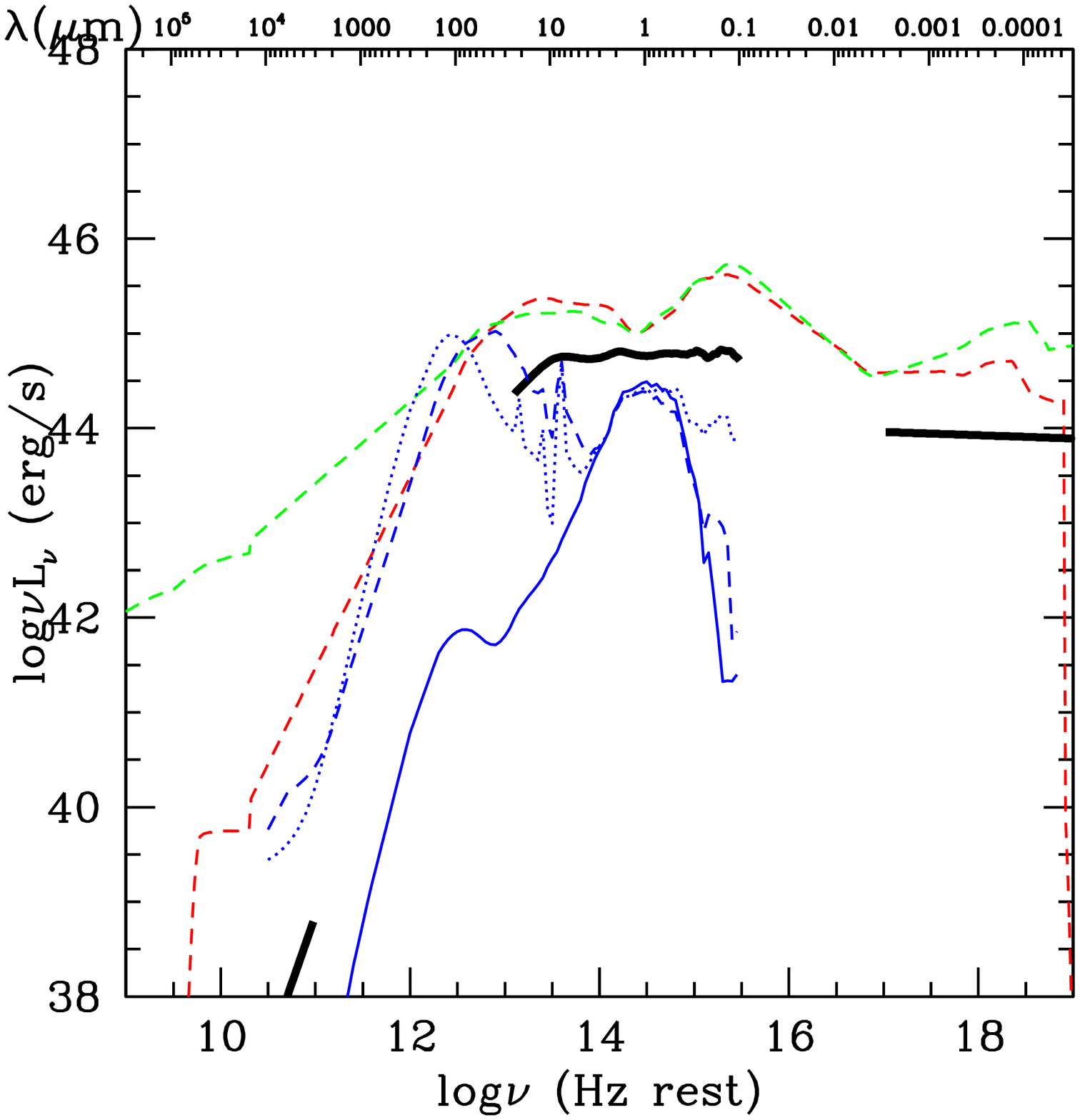}
\includegraphics[angle=0,width=0.5\linewidth]{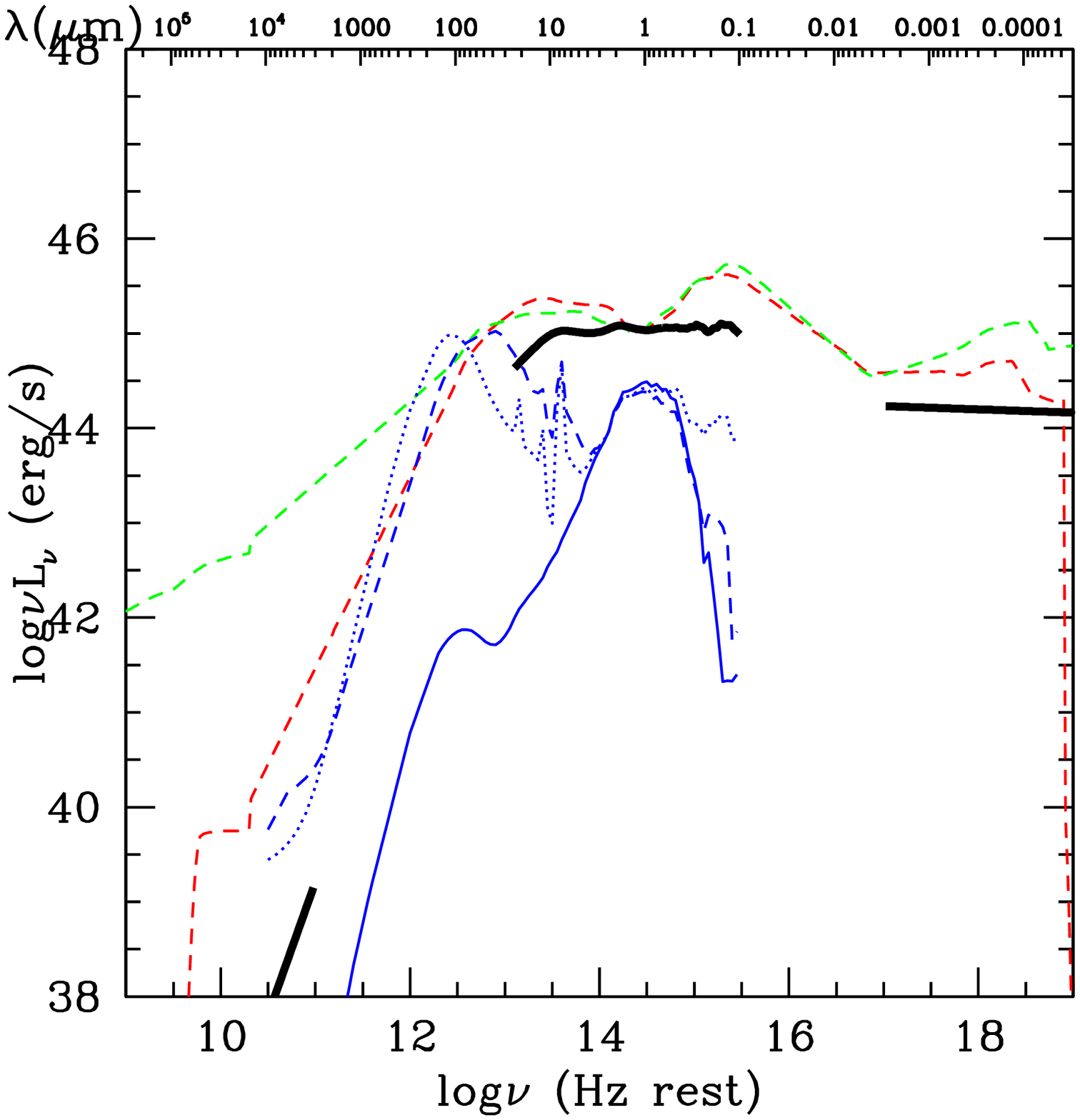}
\caption{The mean SED for all the quasars in XC413 after the
Galactic extinction correction, variability constraints, and broad
emission line correction (black solid line): {\em Left:} arithmetic
mean of the $log\nu L_{\nu}$; {\em Right:} arithmetic mean of the
$log\nu L_{\nu}$ after normalization at 1$\mu$m. The dashed lines
are the mean radio-loud (green) and radio-quiet (red) SEDs from E94.
The blue lines show galaxy templates from Polletta et al. (2007),
normalized to the UKIDSS $L_K^*$ value; blue solid line = elliptical
galaxy with 5~Gyr of age (Ell5), blue dotted line = Spiral galaxy
(Spi4), blue dashed line = a starburst galaxy (NGC6090).
\label{msed}}
\end{figure*}

\begin{figure}
\epsfig{file=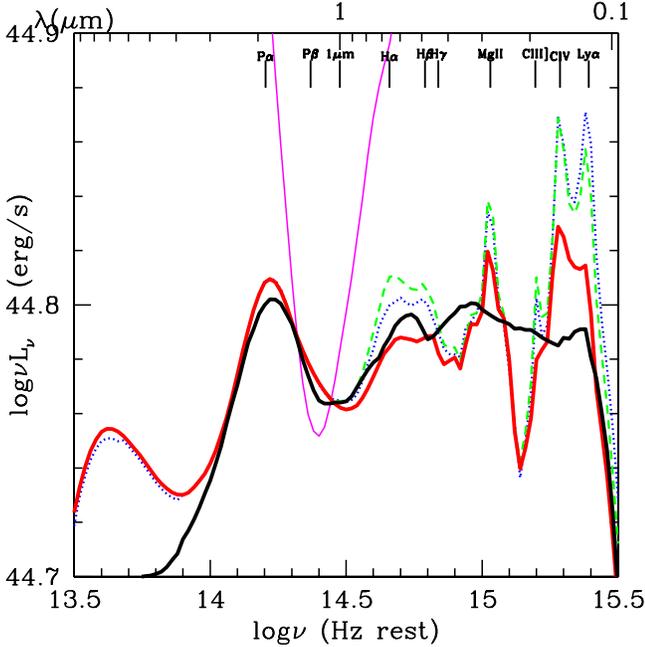, angle=0, width=\linewidth} \caption{Comparison
of the mean IR to UV XC413 SED derived by different methods. (Note
that the y-axis has a much expanded scale compared with
Figure~\ref{egf}). Blue dotted line = `raw' mean SED before
corrections; green dashed line= mean SED after the Galactic
extinction correction and variability restriction. The two solid
lines are the mean SEDs after the Galactic extinction correction,
variability restriction and BEL EW correction: red line = mean SED
by linear interpolation; black line = mean SED by polynomial
fitting. The E94 radio quiet mean SED is shown in magenta solid
line. \label{msedcomp}}
\end{figure}

\begin{figure*}
\includegraphics[angle=0,width=0.5\linewidth]{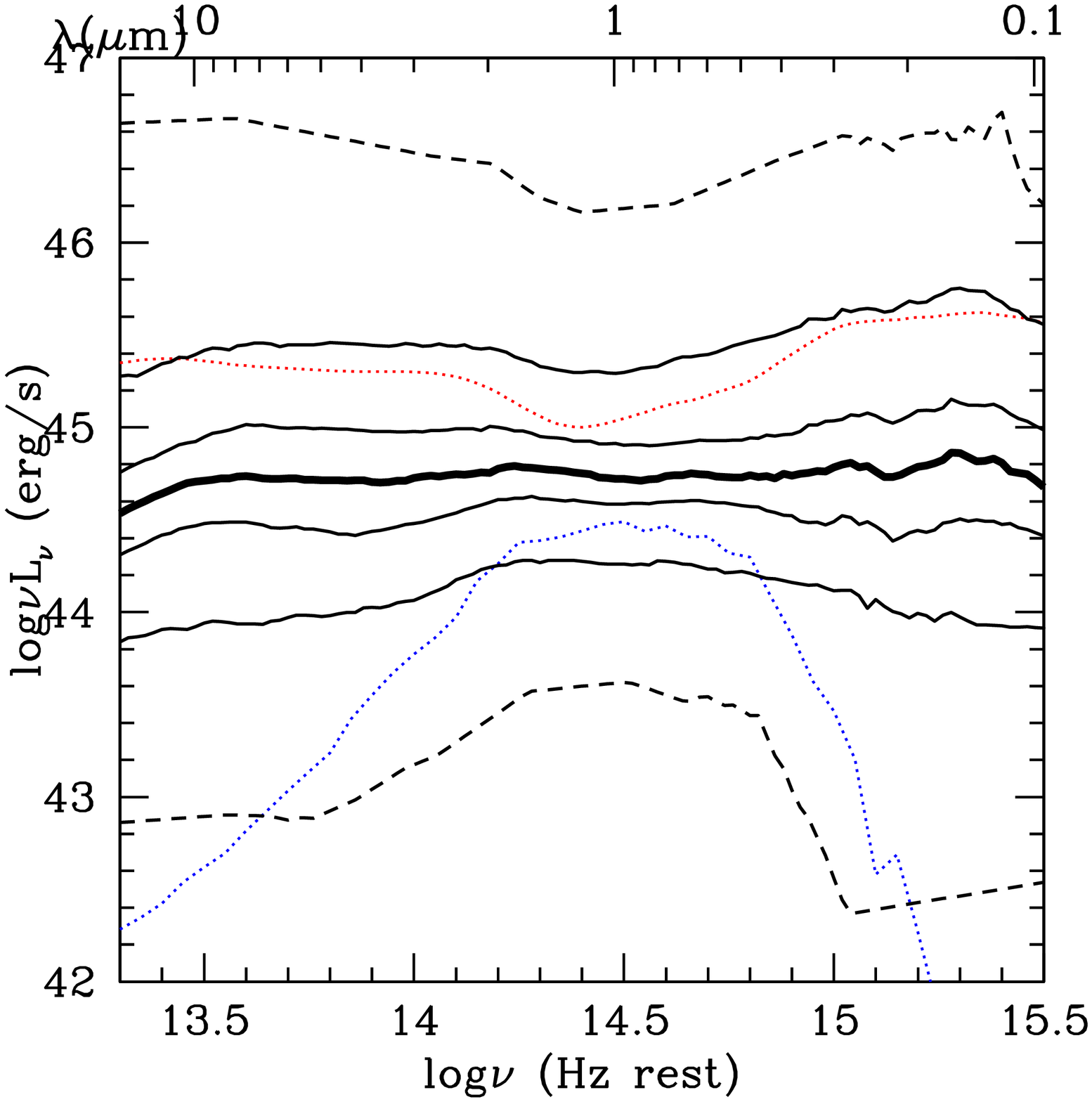}
\includegraphics[angle=0,width=0.5\linewidth]{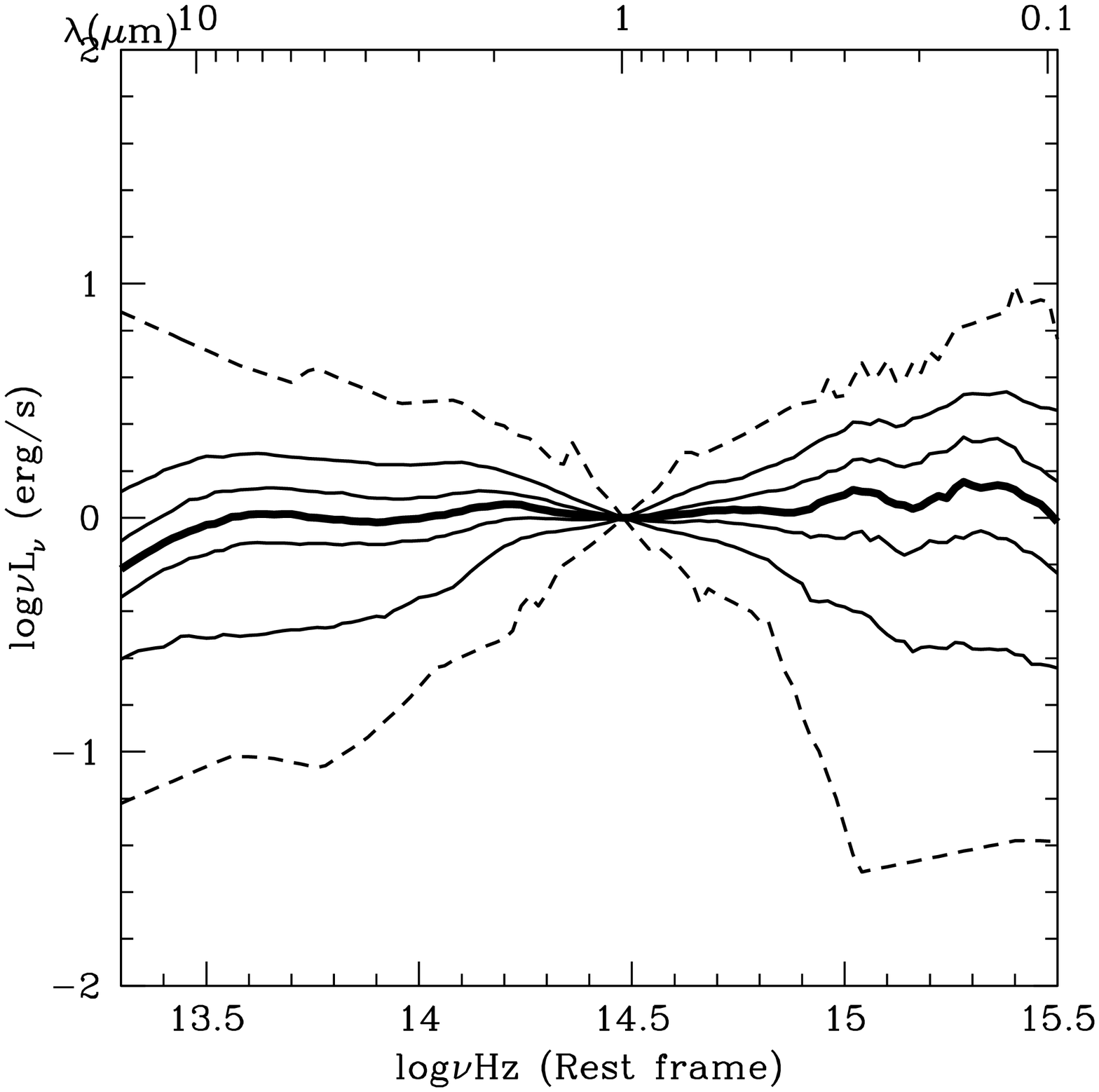}
\caption{The pre-host subtraction median (thick black line) SED and
the 68\%, 90\% (thin solid black lines) and 100\% (dashed black
lines) percentile envelopes in the ultraviolet to infrared
(0.09$\mu$m-24$\mu$m) range:{\em Left:} without normalization; {\em
Right:} normalized at 1$\mu$m. Red dotted line= E94 RQ mean SED.
Blue dotted line=5 Gyr elliptical galaxy template (Ell5).
\label{seddis}}
\end{figure*}

The Galactic extinction, variability and emission line corrections
done in the previous sections are based on reasonable assumptions
without a strong model dependency. We can check the mean and
dispersion of the SEDs after these corrections.

We used both the linearly interpolated SEDs and the SED fitting to
calculate the mean and dispersion of the sample SEDs. We calculated
the arithmetic mean of the $log\nu L_{\nu}$ at each frequency of the
grid. As described above, in the rest frame infrared to ultraviolet
range, both methods are based only on detection (see details in
\S~\ref{s:interpolation}).

The mean SED of the XC413 AGN, after applying all the above
corrections, is shown in Figure~\ref{msed} (left) as a black solid
line, as well as the E94 mean radio-quiet and radio-loud SEDs (red
and green dashed line, respectively). For comparison, we also plot
the mean SED normalized at 1$\mu m$, as for the E94 mean SEDs,
(Figure~\ref{msed} right), to avoid a bias towards high luminosity
sources. The XC413 mean SED is definitely flatter at all wavelengths
than both E94 SEDs.

Three galaxy templates from SWIRE\footnotemark (Polletta et al.
2007) are also shown in the same plot (Figure \ref{msed}): a spiral
galaxy (Spi4), an elliptical galaxy with an age of 5~Gyr, and a
starbust galaxy (NGC6090). These templates are normalized to the
value of $log L_K^*=44.32$ ($M_K^*=-23$) from the UKIDSS Ultra Deep
Survey (Cirasuolo et al. 2007). The XC413 mean SED could be flat
because affected by contamination from a strong host galaxy
component. \footnotetext{The 16 galaxy templates in the ``SWIRE
Template Library'' (Polletta et al. 2007) include: 3 elliptical
galaxy templates ``Ell2'', ``Ell5'', ``Ell13'' representing
elliptical galaxy of age 2~Gyr, 5~Gyr and 13~Gyr respectively; 7
spiral galaxy templates ``S0'', ``Sa'', ``Sb'', ``Sc'', ``Sd'',
``Sdm'', ``Spi4''; and 6 starburst galaxy templates ``NGC6090'',
``M82'', ``Arp220'', ``IRAS20551-4250'', ``IRAS22491-1808'',
``NGC6240''.}

We compare the infrared to ultraviolet mean SED before and after the
corrections for Galactic reddening, variability and BEL EW on a much
expanded scale in Figure~\ref{msedcomp}. The differences are
strongly correlated with the location of the BELs, meaning that that
the BEL has been properly corrected in at least 8 out of 9 (90\%)
lines. Even though the EW distribution (Figure 12) does not show a
peculiar behavior, the CIV emission line correction, as derived from
SDSS, is not sufficient in our sample ($\sim$0.02 dex higher than
continuum). The polynomial fit makes the SED shape smoother compared
to linear interpolation, however the difference between the two
methods is less than 0.02 dex in the 0.1--1 $\mu m$ range.

Concentrating on the UV to IR (0.1--10$\mu$m) range, it is clear
from Figure~\ref{msed} that the mean SED of the XC413 sample is
quite flat, and lacks the clear 1$\mu$m inflection point between the
UV and near-IR bumps seen in E94. E94 found that host galaxy
corrections were significant even in the most luminous quasars in
their sample\footnotemark.
\footnotetext{In retrospect, this result, like those of McLeod et
al. (1994) and Eskridge et al. (1995), was an early hint of the
correlation between host bulge mass and central black hole mass
(Magorrian et al. 1998).}

The dispersion of the SEDs, computed using the median SED and the
68, 90 and 100 percentile contours on each side of the median are
shown in Figure~\ref{seddis}, together with the E94 SED.  The upper
90th percentile SED shows a E94 like shape, while the lower 90th
percentile SED is close to the galaxy templates shown in
Figure~\ref{seddis}. Typically X-ray selected AGN samples, such as
XMM-COSMOS, include more sources with low AGN to host galaxy
contrast, compared to optically selected samples, and so are more
affected by the host galaxy contribution. Due to the bias of the
optical color selection, optically selected samples tend to select
sources with clear big blue bumps, thus missing the many quasars
with large host contributions (compare to Richards et al. 2006 and
Luo et al. 2010).

\subsection{Host Galaxy Correction}
\label{s:host}

\begin{figure}
\epsfig{file=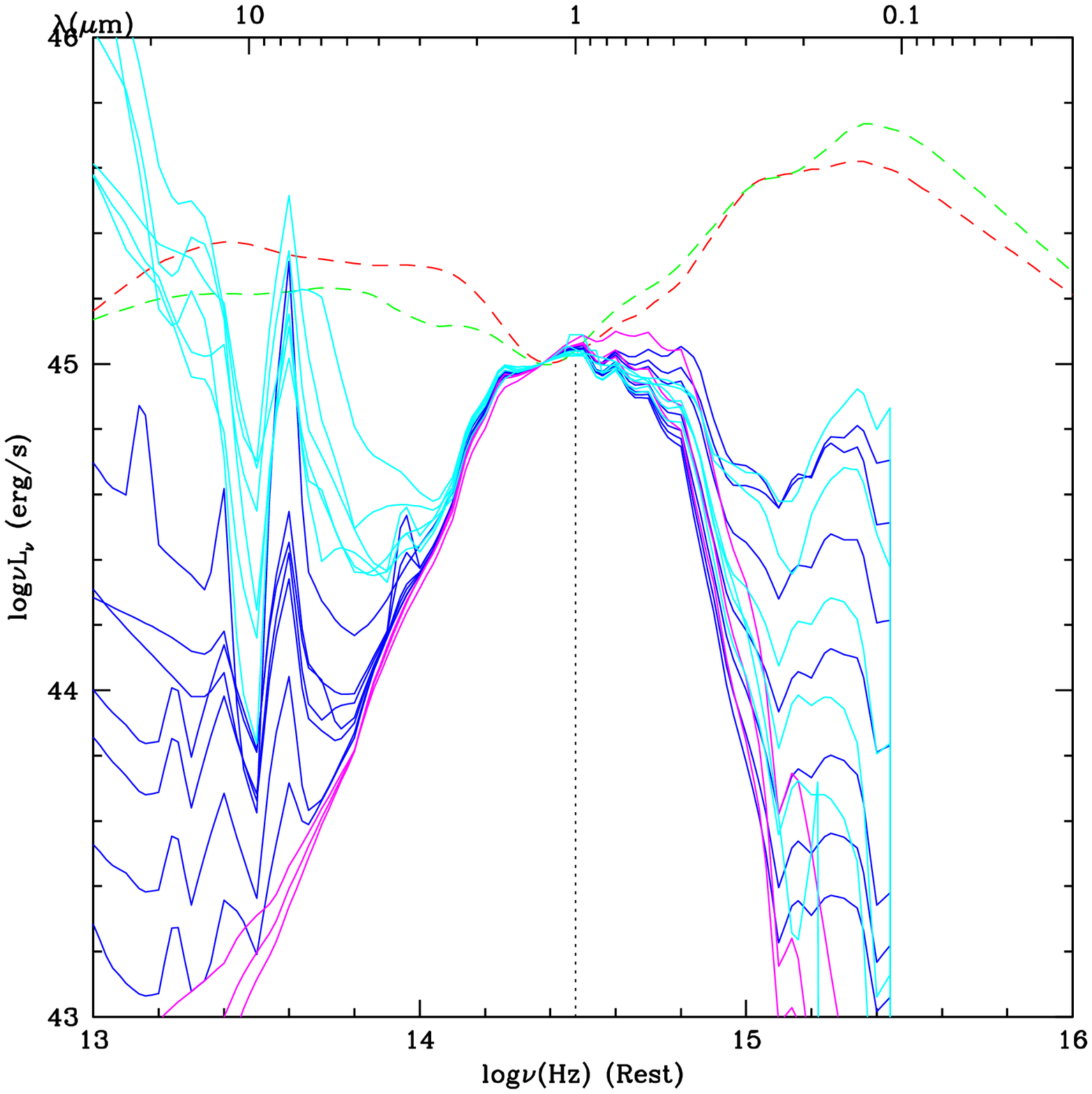, angle=0, width=\linewidth} \caption{ SED of 16
galaxy template from SWIRE template library (Polletta et al. 2007)
normalized at rest frame J band are shown as solid line (blue=spiral
galaxy: S0, Sa, Sb, Sc, Sd, Sdm and Spi4; magenta=elliptical galaxy
with 2~Gyr, 5~Gyr and 13~Gyr of age; and cyan=starburst galaxy:
NGC6090, M82, Arp220, IRAS20551-4250, IRAS22491-1808, NGC6240). The
E94 mean SED are shown as dashed lines (red as radio quiet and green
as radio loud). \label{galtemp}}
\end{figure}

\begin{figure}
\epsfig{file=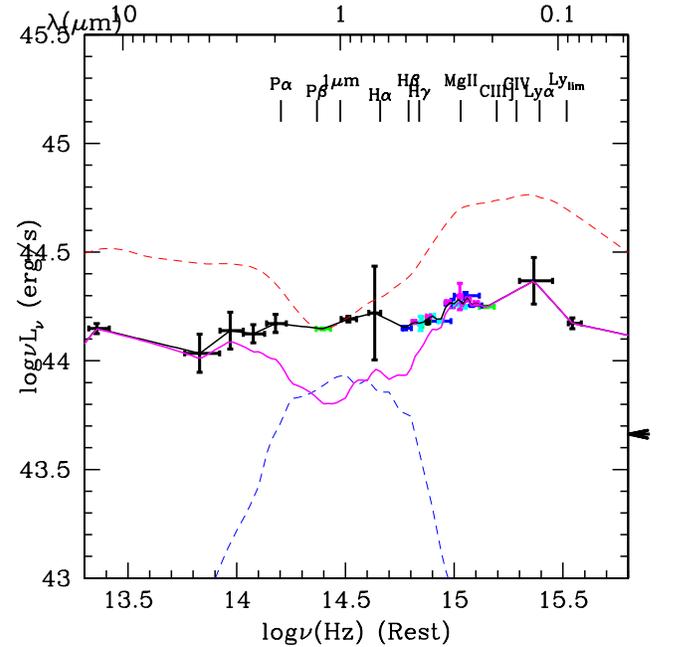, angle=0, width=\linewidth} \caption{SED of
source COSMOS\_J150.50924+2.699418 (XID=86) at z=0.794 (i-band
absolute magnitude -22.4). The mass of the black hole is $log
M_{BH}=8.24$. The black solid line show the SED as the interpolation
of the observed data. The magenta solid line shows the SED after
host galaxy correction. Red dashed line= E94 RQ mean SED. Blue
dashed line=host galaxy template (5Gyr elliptical galaxy) normalized
at $L_{J,Gal}$ calculated from the Marconi \& Hunt (2003) scaling
relationship adding the evolutionary term. The points are color
coded as in Figure~\ref{vareg}. \label{eghost}}
\end{figure}

\begin{figure*}
\includegraphics[angle=0,width=0.5\linewidth]{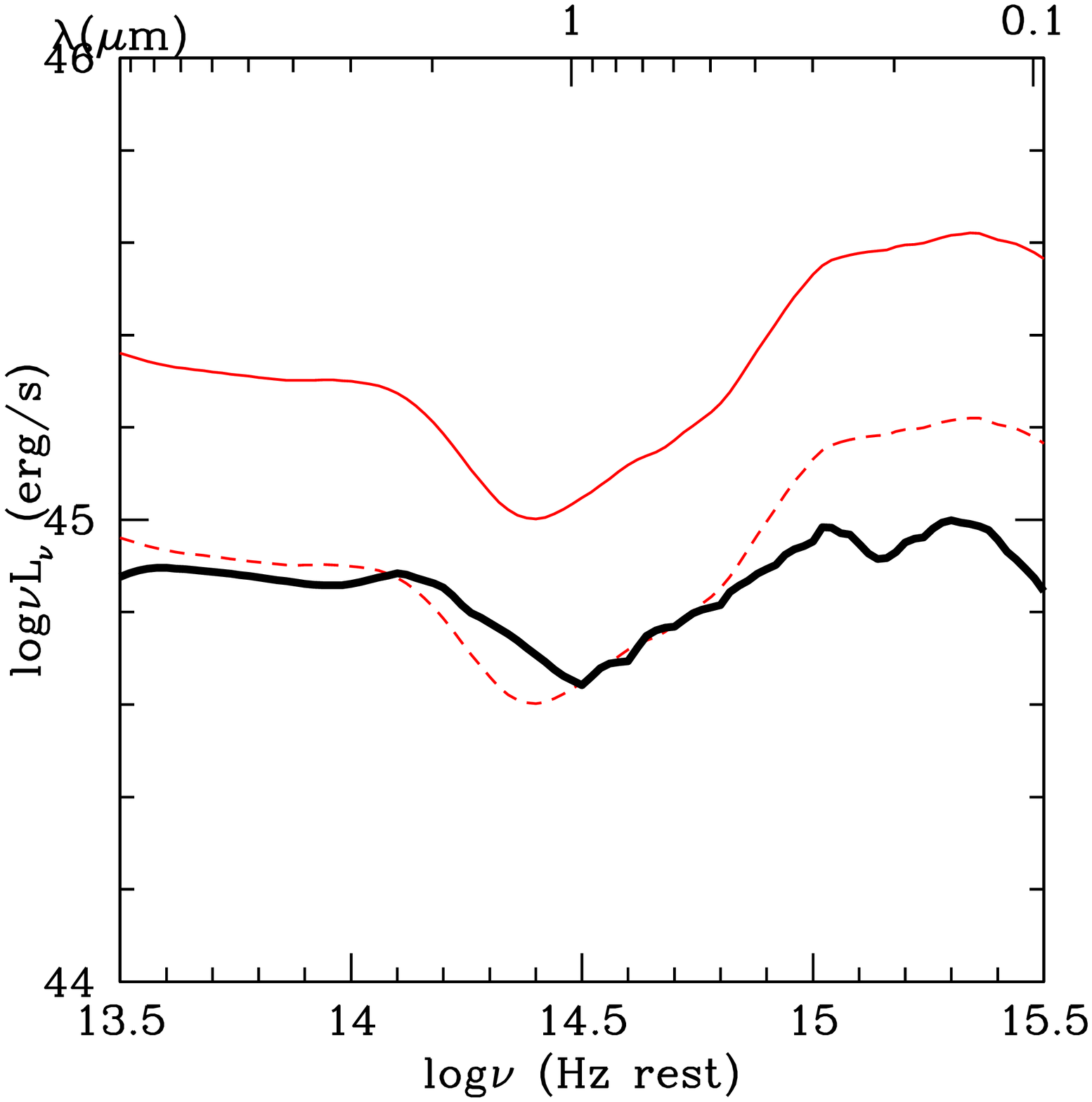}
\includegraphics[angle=0,width=0.5\linewidth]{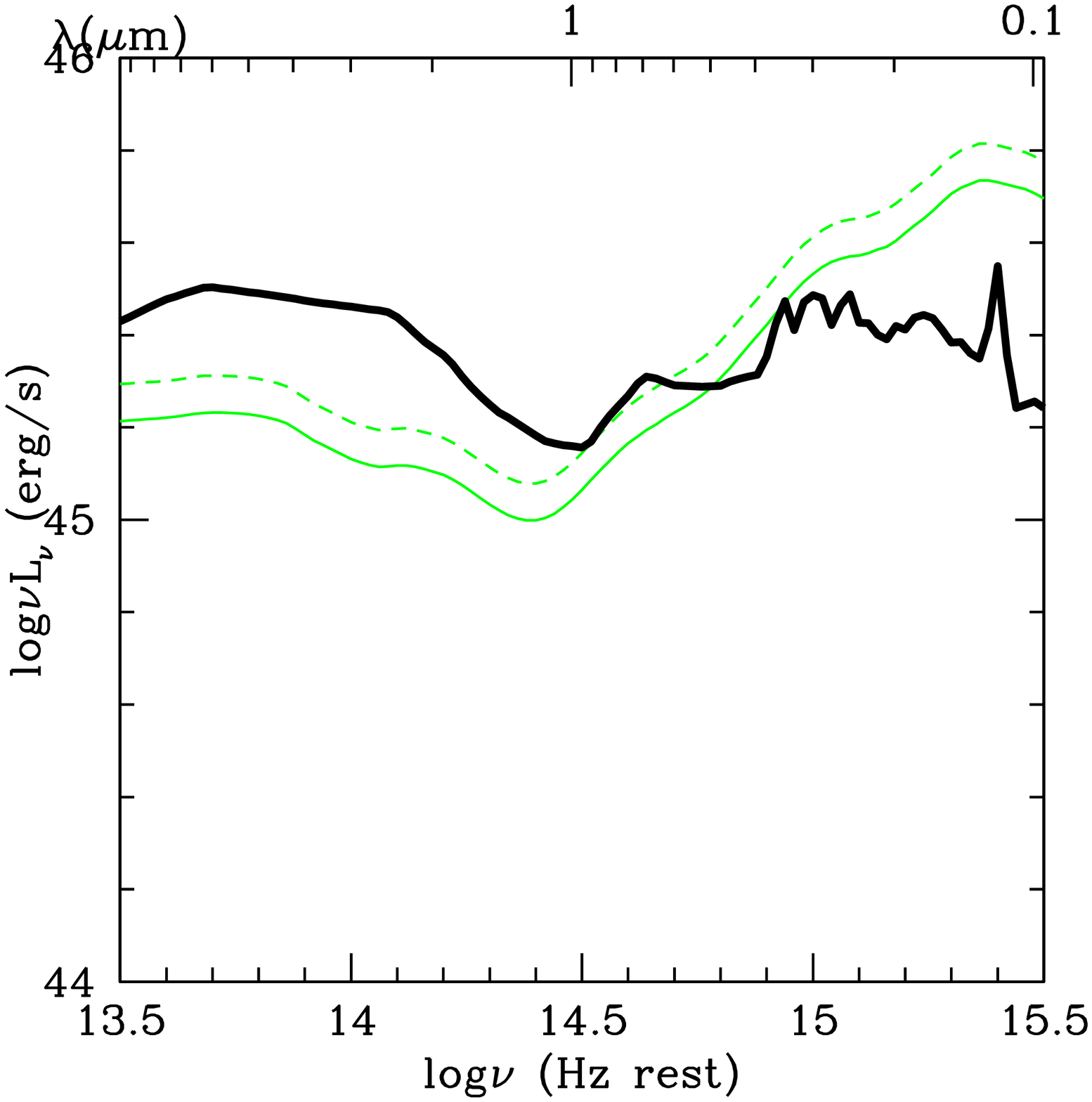}
\caption{{\em Left:} The mean IR to UV SED for 199 radio quiet (RQ)
sources in our sample after all corrections including host galaxy
subtraction (black solid line). The red lines are the E94
radio-quiet mean SED (the dashed line is normalized to the XMM
sample). {\em Right:} The mean IR to UV SED for 4 radio loud (RL)
quasars after all corrections (black solid line). The green lines
are the E94 radio-loud mean SED (dashed is normalized to the XMM
sample). \label{msedhc}}
\end{figure*}

\begin{figure*}
\includegraphics[angle=0,width=0.5\linewidth]{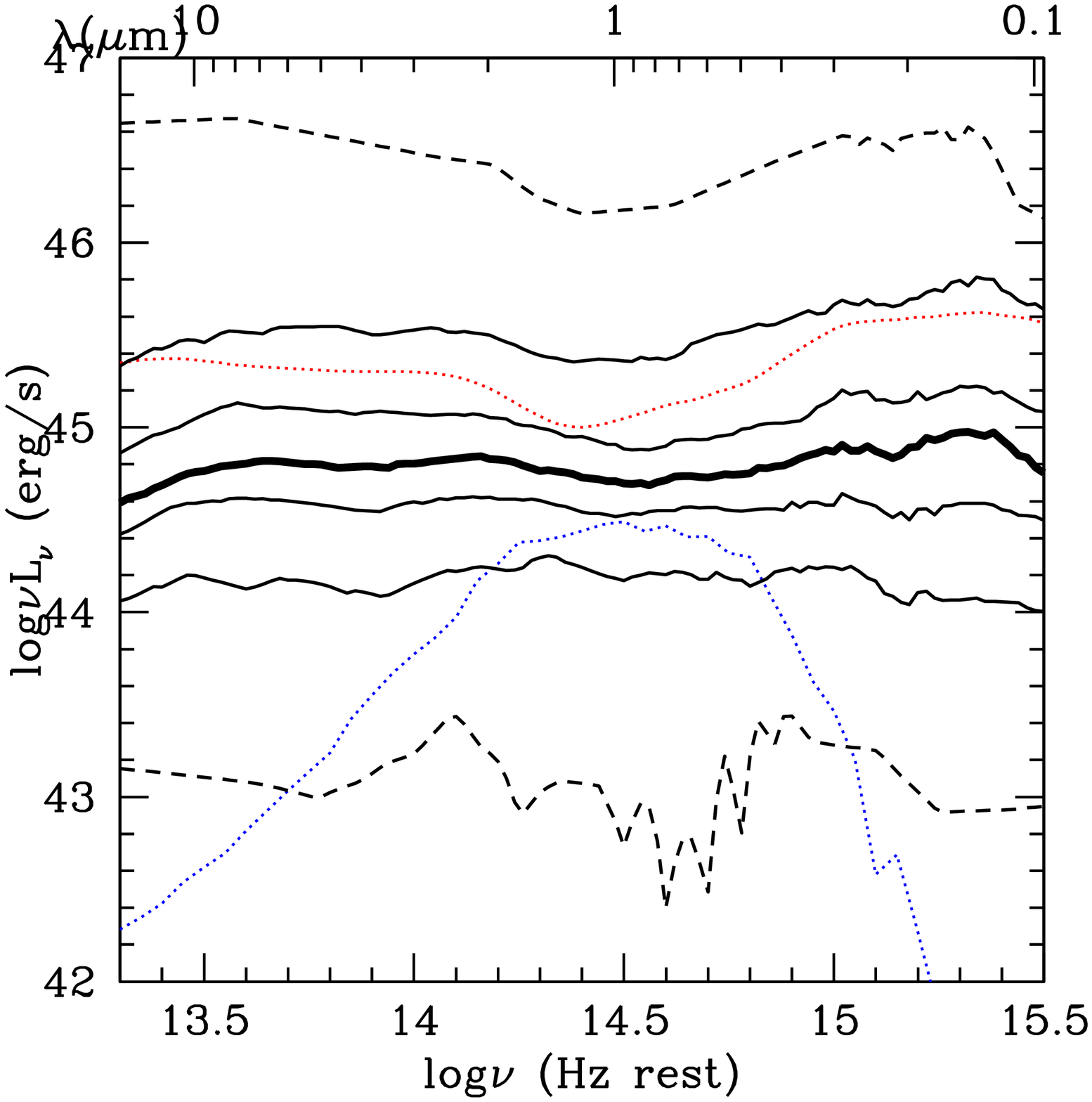}
\includegraphics[angle=0,width=0.5\linewidth]{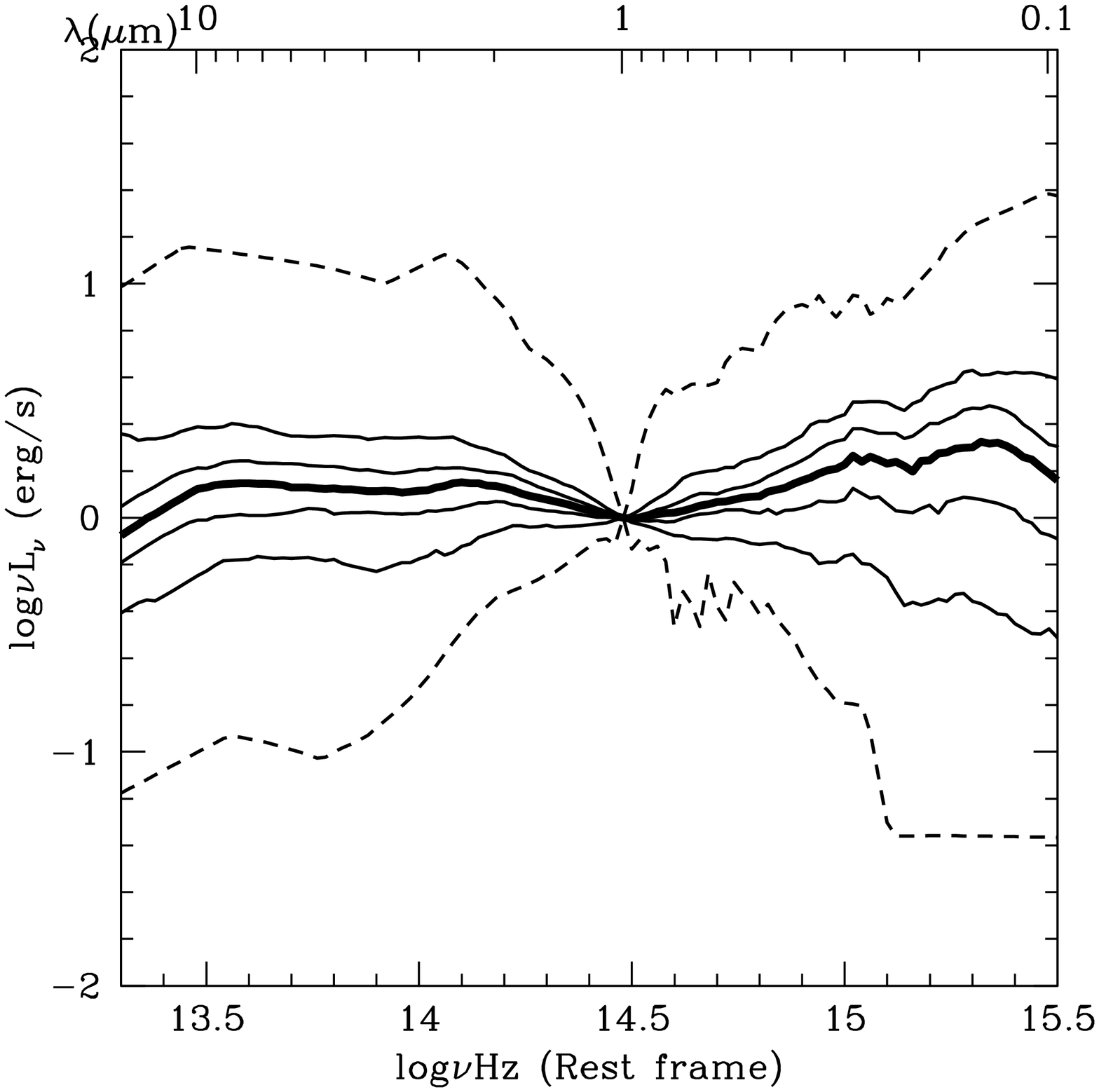}
\caption{The median (thick black line) SED and the 68\%, 90\% (thin
solid black lines) and 100\% (dashed black lines) percentile
envelopes in the ultraviolet to infrared (0.9$\mu$m-24$\mu$m) range
after all correction including the host galaxy subtraction: {\em
Left:} before normalization; {\em Right:} normalized at 1$\mu$m. Red
line= E94 SED.  Blue dotted line=5 Gyr elliptical galaxy template
(Ell5). \label{seddishc}}
\end{figure*}

\begin{figure} \epsfig{file=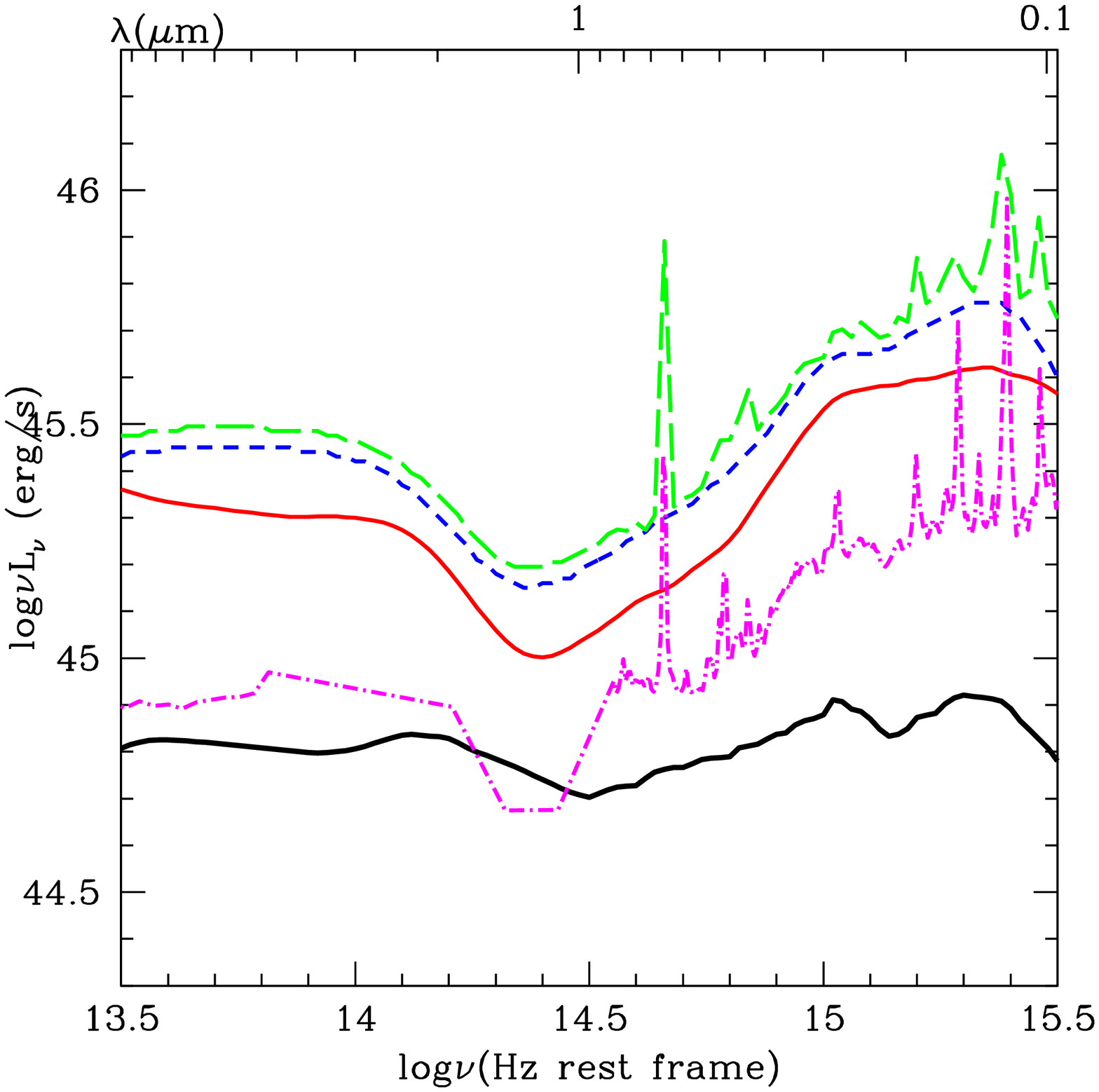, angle=0,
width=\linewidth} \caption{ The comparison of the mean
host-corrected SED for the 199 radio-quiet quasar of XC203 (black
solid line) with previous studies. The red solid line shows the E94
radio quiet mean SED. The blue short dashed line shows the Richards
et al. (2006) mean SED. The green long dashed line shows the Hopkins
et al. (2007) AGN SED template. The magenta dot-dashed line shows
the Shang et al. (2011) mean SED. \label{SEDtmpcmp}}
\end{figure}

Jahnke et al. (in prep.) uses the COSMOS high-resolution HST/ACS
F814W image (Koekemoer et al. 2007) to estimate the host galaxy
luminosity and AGN luminosity for XMM-COSMOS sources at $z<1$. We
could use their results by normalizing a galaxy template to the
galaxy luminosity at the observed frequency and subtracting the
galaxy contribution from the observed SED. However, this method does
not work well for our sample, for the following reasons:

\begin{enumerate}
\item We have to assume the host galaxy type in which the AGN is harbored.
The F814W band would be at $\sim3000$\AA\ rest frame for a typical
XC413 quasar at redshift $z=1.5$, which lies on the steeply falling
blue side of the galaxy template, so that a small error in template
slope (or, effectively, in the age of the youngest population in the
host) would lead to a large error in the normalization
(Figure~\ref{galtemp}). In the XMM-COSMOS sample, this issue leads
to a severe over-subtraction problem in some cases. For a 5~Gyr old
elliptical galaxy template, 9 out of the 89 sources with host
magnitude estimation are over-subtracted (10\%). For spirals and
star-burst galaxies, this method completely fails, as the
over-subtraction fraction is $>$80\%.

\item Direct imaging works only for the 89 sources at low redshift ($z<1$) for
which the HST F814W image gives a good host galaxy magnitude estimates (Jahnke
et al. in prep.). For the remaining 324 sources in the sample, we cannot perform
a host galaxy correction with this method.
\end{enumerate}

Estimates of host luminosity have been made using other techniques:

(1) Vanden Berk et al. (2006) use the eigenspectrum decomposition
technique to give a host to AGN relationship in r band, which can be
transformed to luminosity as follows (Richards et al. 2006,
hereinafter R06).
\begin{equation}
log(L_{r,Gal})=0.87log(L_{r,AGN})+2.887-log\lambda_E
\end{equation}
where $\lambda_E$ is the Eddington ratio.

(2) Marconi \& Hunt (2003) analyze two-dimensional images of nearby
galaxies with a black hole mass determined with direct gas
kinematics or stellar dynamics (Tremaine et al. 2002) in the local
universe. Their results can be transformed as below:
\begin{eqnarray}
log(L_{B,Gal})&=&0.84log(L_{bol})+3.914-0.84log\lambda_E\\
log(L_{J,Gal})&=&0.88log(L_{bol})+3.545-0.88log\lambda_E\\
log(L_{H,Gal})&=&0.86log(L_{bol})+4.530-0.86log\lambda_E\\
log(L_{K,Gal})&=&0.88log(L_{bol})+3.577-0.88log\lambda_E
\end{eqnarray}

The above relation is derived for the local universe. However there
is evidence that the black hole mass and host luminosity
relationship evolves with redshift (Peng et al. 2006a,b; Ho 2007;
Decarli et al. 2010; Merloni et al. 2010; Bennert et al. 2010,
2011). Allowing for selection effects, the evolutionary trend is
$M_{BH}/L_{sph}\propto (1+z)^{1.4\pm0.2}$ (Bennert et al. 2010,
2011). As most XC413 quasars are at high redshift, we need to
include this effect. We added an additional redshift term in the
above equations (4--7), giving the following equations:

\begin{eqnarray}
log(L_{B,Gal})&=&0.84log(L_{bol})+3.914-0.84log\lambda_E\nonumber\\
& &-1.18log(1+z)\\
log(L_{J,Gal})&=&0.88log(L_{bol})+3.545-0.88log\lambda_E\nonumber\\
& &-1.23log(1+z)\\
log(L_{H,Gal})&=&0.86log(L_{bol})+4.530-0.86log\lambda_E\nonumber\\
& &-1.20log(1+z)\\
log(L_{K,Gal})&=&0.88log(L_{bol})+3.577-0.88log\lambda_E\nonumber\\
& &-1.23log(1+z)
\end{eqnarray}

We choose the rest frame J band luminosity $L_{J,gal}$, because this
is the band closest to $1\mu$m, where the galaxy contributes
strongly. The rest frame J band is also located on the flat part of
the host templates, so the normalization is insensitive to
uncertainties in the template slope, in contrast to the observed
F814W band for quasars at redshift 1--2.

Note that, in these formulae, $L_{bol}$ has the same coefficient as
$\lambda_E$, thus the host galaxy luminosity is physically a
function of the black hole mass only (see Marconi \& Hunt 2003 for
details). To apply these formulae, black hole mass estimates are
needed. The XC413 quasars have 206 published black hole mass
measurements (Trump et al. 2009b; Merloni et al. 2010), which are
based on the scaling relationship between BEL FWHM and black hole
mass (Vestergaard 2004). For the quasars with only zCOSMOS spectra,
the black hole mass was estimated for only the ones with MgII lines
in the spectra (Merloni et al. 2010),  using the calibration of
McLure \& Jarvis (2002). For the rest of the sample, the BELs are
located at the edge of the spectrum, so that reliable black hole
mass estimates cannot be made. We calculated bolometric luminosities
by directly integrating the SED from 24$\mu$m to 40 $keV$ (see Paper
2, Hao et al. 2012 in prep.). The rest-frame J band galaxy
luminosity is then calculated using equation (9) for the 206 objects
with black hole masses, hereinafter the ``XC206'' sample.

We need to assume a host galaxy SED to do the corrections. We have
no information on the galaxy type of the host for most of the
quasars, even with the help of the HST ACS images, as the bulk of
the XMM-COSMOS quasars are at $z>1$. We checked the 16 galaxy
templates from the SWIRE template library (Polletta et al. 2007). We
normalized all the template to the rest frame J band in
Figure~\ref{galtemp}. We find that none of our quasars show obvious
PAH feature in the mid-infrared, which is a prominent feature of
starburst galaxy spectra. This could, however, be due to the gap in
photometry coverage between the Spitzer bands. From the Spitzer
photometry alone, it is difficult to detect such features. Starburst
galaxy templates with prominent PAH feature are not suitable for
host correction, though, as the PAH feature could exceed the
observed (interpolated) SED. The elliptical and spiral galaxies have
similar SED shapes from 0.4--4$\mu$m, with a dispersion less than
0.1 dex. This wavelength range is where the majority of the host
galaxy contribution lies. We checked all spiral or elliptical galaxy
templates and the correcting results are similar to each other.
Therefore, for the purpose of host correction, it makes no
difference which templates to choose.

We used a 5 Gyr old elliptical galaxy template (Polletta et al.
2007) normalized to the rest-frame J band galaxy luminosity, and
subtracted the galaxy contribution from the observed SED. An example
is shown in Figure~\ref{eghost}. In this case, the observed SED is
rather flat. After subtracting a scaled host contribution (blue
dashed line), the corrected SED (magenta solid line) shows a shape
close to that of E94 (red dashed line).

This subtraction is successful in 203 cases in XC206. In just 3
cases (1.46\%) we have an over-subtraction problem, and 2 out of the
3 are over-subtracted by less than 0.1 dex.  By over-subtraction, we
mean the normalized host galaxy SED exceeds the observed SED at
certain wavelength range (usually at $\sim 1 \mu$m, where the galaxy
contributes the most). If we use the local scaling relationship
directly, i.e. without the redshift correction, there would be 34
quasars in which the estimated host flux exceeds the observed flux,
leading to larger over-subtraction problem. This suggests that the
evolutionary term is needed, and gives some support to the idea that
the M-$\sigma$ relation evolves. The over-subtraction could be due
either to dispersion or evolution of the scaling relationship. On
the other hand, under-subtraction also happens in the correction,
which is more difficult to classify. By under-subtraction, we mean
that the correction leaves a flat SED at around 1$\mu$m.

The host correction based on the scaling relationship we applied
here is probably not the most accurate, but is the best available at
present. We applied this host galaxy correction for these 203
quasars, hereinafter ``XC203''. The XC203 quasars are plotted as
green hexagons in Figure \ref{Lxz}--\ref{hrz}. In this sub-sample,
four\footnotemark are radio-loud according to the criteria discussed
in \S \ref{s:radio}.
\footnotetext{XID=5230, 5275, 5517, 54541.}
We calculated the mean SED and dispersion SED as the previous section.

The mean and dispersion of the XC203 SEDs after all the corrections,
including galaxy correction, resembles the E94 mean SED (Figure
\ref{msedhc} and \ref{seddishc}).

In Figure~\ref{SEDtmpcmp}, we compared the mean XC203 SED with other
mean SEDs: E94 radio quiet, R06, Hopkins et al. (2007) and the Shang
et al. (2011). The R06 SEDs used a ``gap-repair'' technique, that
replace the missing values with normalized the E94 mean SED to the
adjacent photometry bands. As the R06 has limited coverage in near
infrared J H K bands (J:40/259, H:35/259, K:42/259), the mean SED is
therefore, by construction, similar to the E94 radio quiet mean SED.
The Hopkins et al. (2007) just combined the R06 mean SED template
with the composite quasar SED (Vanden Berk et al. 2001), and so
their SED template has a similar shape as R06. The Shang et al.
(2011) mean SED is calculated using nearby bright quasars, which
have the same selection bias as E94 and so finds a similar shape as
E94. Compared to all these SED templates, the XC203 mean SED is
flatter due to possible excess host contribution, not corrected
because of the dispersion in the scaling relationship itself.

\subsection{The Variety of Type 1 AGN SEDs in COSMOS}
\label{s:variety}

\begin{figure*}
\includegraphics[angle=0,width=0.5\textwidth]{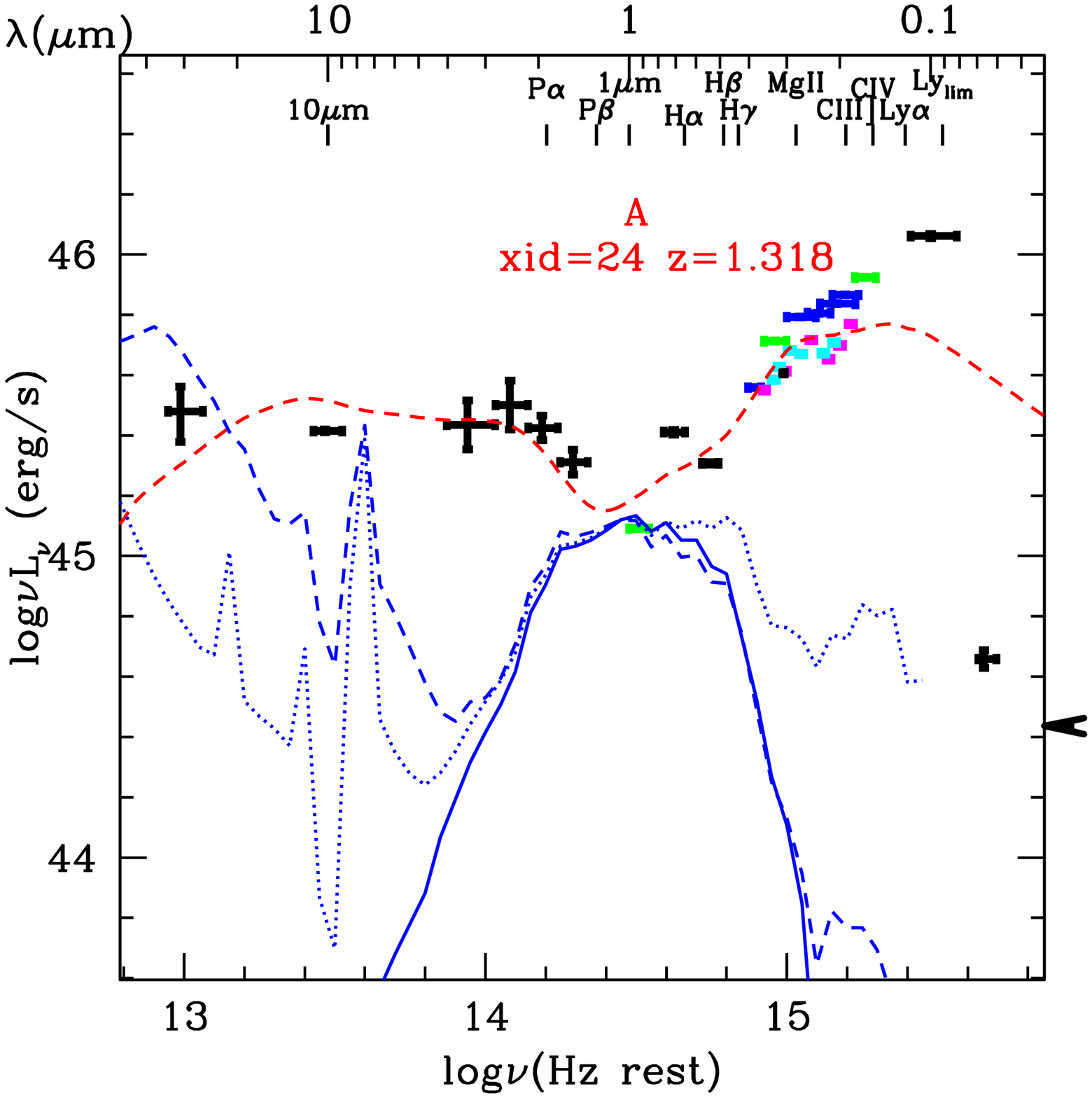}
\includegraphics[angle=0,width=0.5\textwidth]{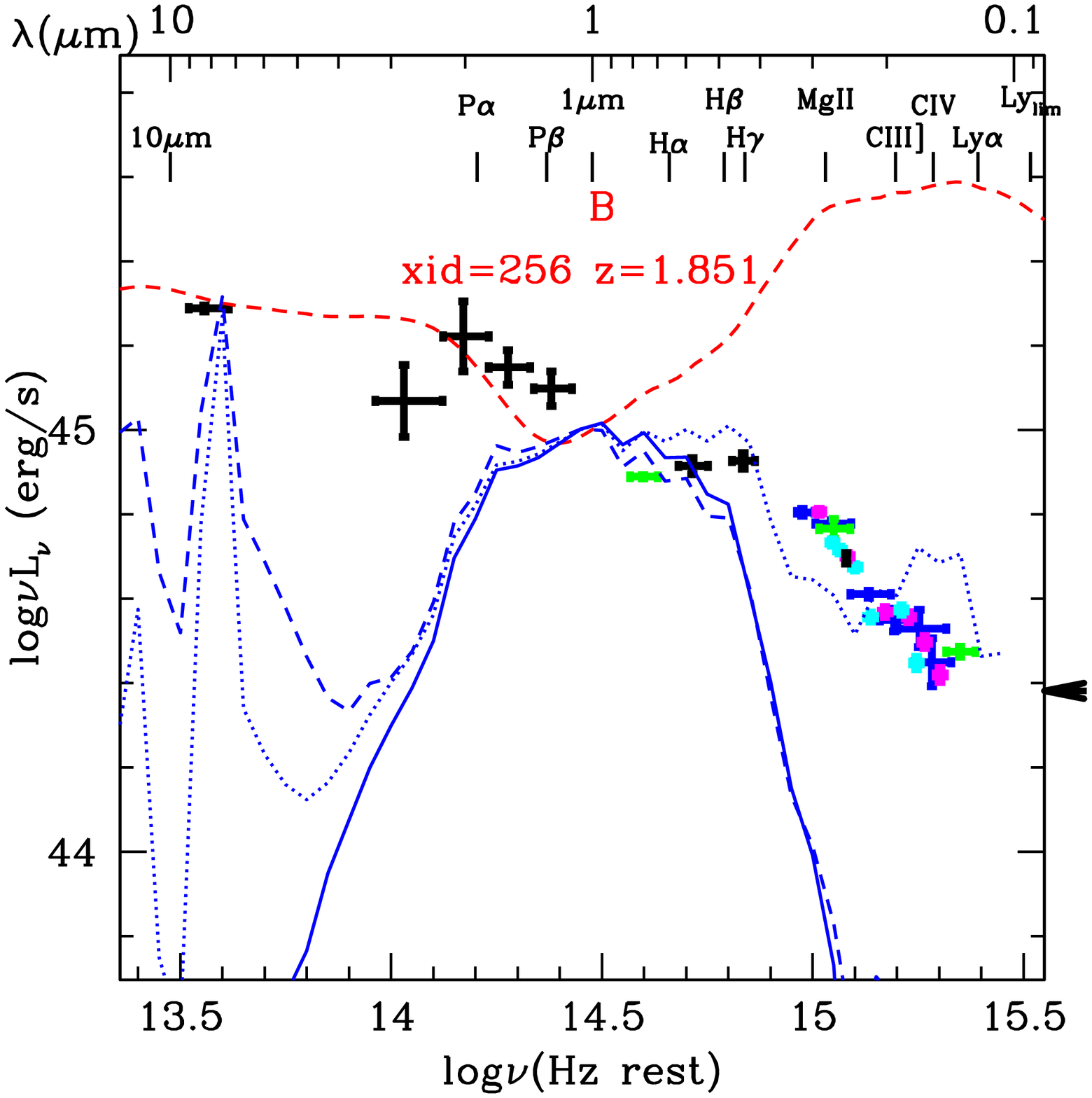}
\includegraphics[angle=0,width=0.5\textwidth]{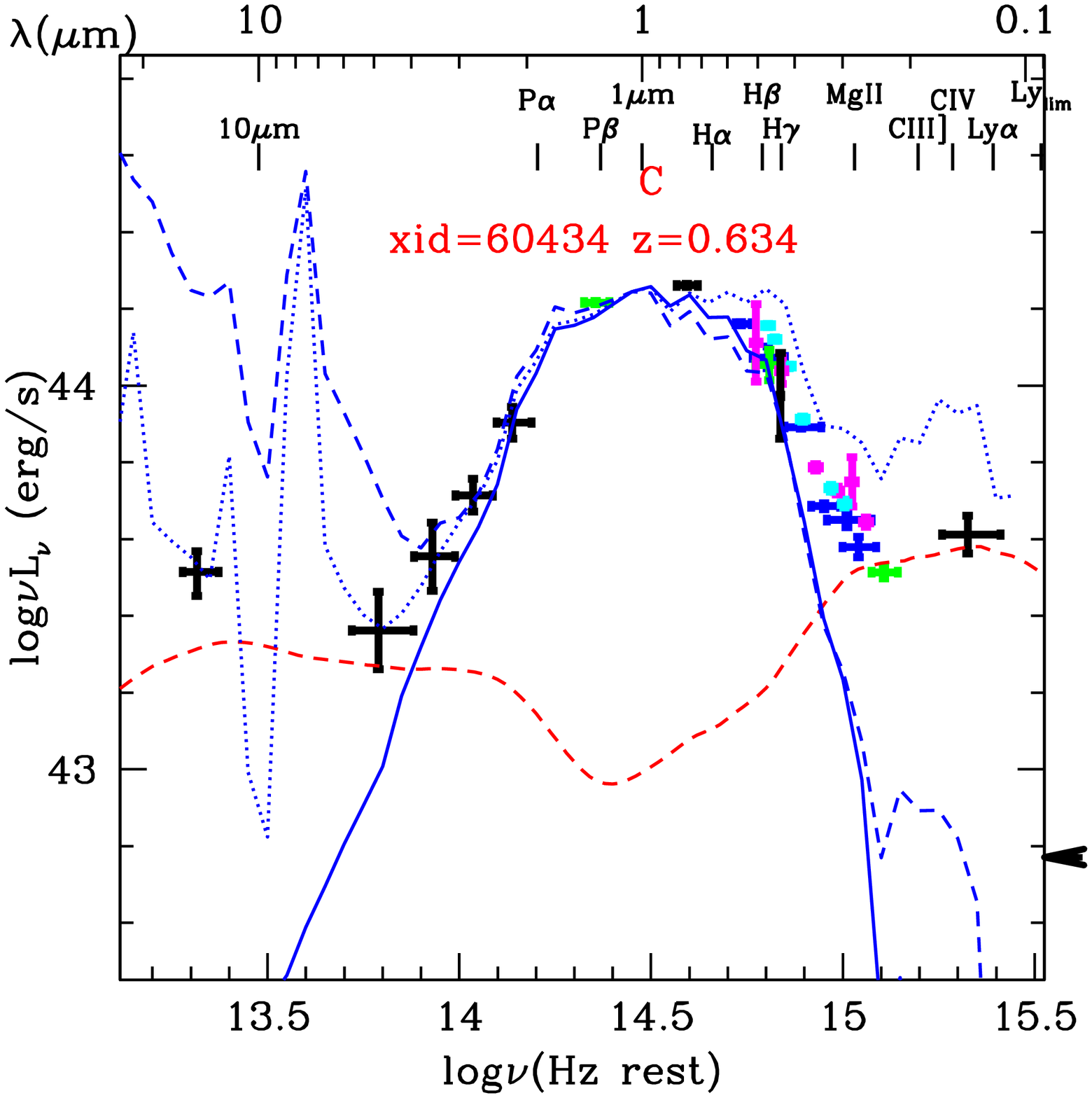}
\includegraphics[angle=0,width=0.5\textwidth]{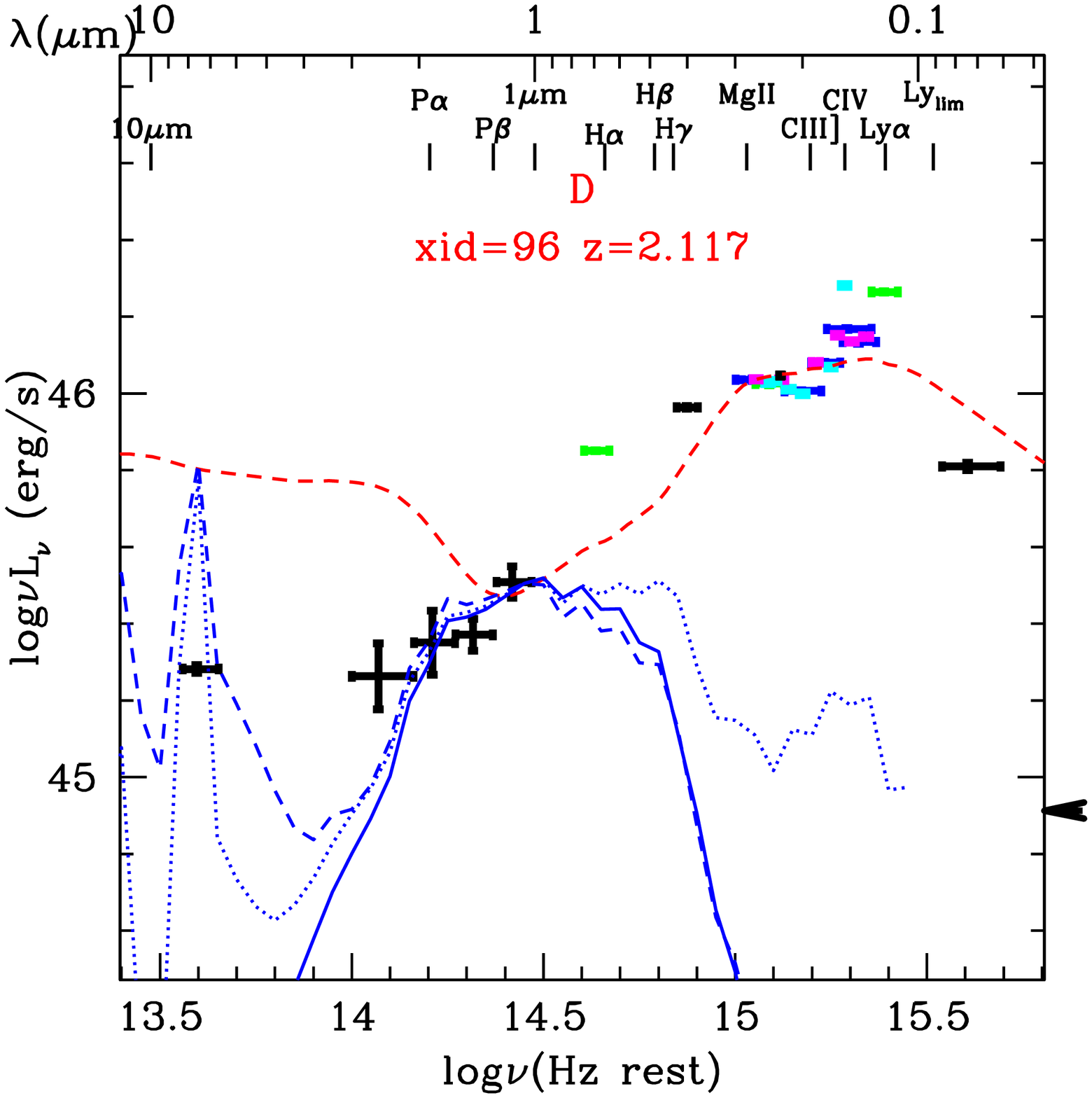}
\caption{Extreme examples of SEDs: {\em top left:} A. a close analog
of the E94 mean SED; {\em top right:} B. no big blue bump, probably
due to reddening and/or a strong galaxy component; {\em bottom
left:} C. a probable galaxy dominated AGN. ; {\em bottom right:} D.
strong big blue bump but no 1~$\mu$m inflection due to a weak
near-IR bump. The data points and the E94 RQ SED are shown as in
Figure~\ref{vareg}. The galaxy templates are shown as in
Figure~\ref{msed} normalized at $1\mu m$.\label{extremeseds}}
\end{figure*}

Simply using the dispersion in the SEDs (Figure~\ref{seddis}), does not give a
full picture of the variety of SEDs in the XC413 sample. To
illustrate the variety of type~1 AGN SEDs found in the XMM-COSMOS sample, we
have selected four examples that span the range of properties
(Figure~\ref{extremeseds}, A, B, C, D).  The four sources are also marked in
Figure \ref{Lxz}--\ref{hrz}.  These four examples illustrate:

\begin{enumerate}
\item {\em A. `Normal' SED}, with photometry points lying strikingly close to
the E94 radio quiet mean SED (red dashed line), except for the few that are
clearly affected by the broad emission lines H$\alpha$, CIV and Ly$\alpha$, and
the point beyond the Lyman limit.

\item {\em B. 'Reddened' SED}, with an optical/UV SED too blue to be a host
galaxy, but dropping rapidly toward the UV, an effect quite likely caused by
reddening. The $\sim$1~dex drop in the u-band compared to the E94 RQ mean would
correspond to a fairly modest extinction of E(B-V)=0.18, for an SMC extinction
curve (Gordon et al. 2003).

\item {\em C. `Host dominated' SED}, matching closely to galaxy templates. The
observed SED could be fitted as a strong galaxy plus a faint AGN
component. About 10\% of the quasars in the sample have host contribution of
more than 85\% at $\sim 1\mu$m.

\item {\em D. `Hot-dust-poor' SED}, lacking a 1~$\mu$m inflection point, but
with a normal E94-like strong big blue bump. About 10\% of the quasars in the
sample belong to this category. These quasars have been discussed in detail in
Hao et al. (2010a).

\end{enumerate}

As the variety of the SED shapes in the sample is a continuous
distribution, the fraction of the sources of the above four types
depend on how we define the selection criteria for each type.
Besides, for most of the sources, the SED shape can be explained by
combining two or more of the above four types. So the fraction of
the sources in each type can not be easily estimated.

Three of these four types have been seen before in AGN at lower
redshift, but similar luminosity. The early 2-10~keV sky surveys
(e.g. Ariel~V, Cooke et al. 1978, HEAO-1 A2, Piccinotti et al. 1982)
produced samples of a few dozen AGN that have similar SED
characteristics. Ward et al. (1987) compiled U-band to IRAS
100~$\mu$m SEDs for the `Piccinotti' AGN and divided them into three
types based on their $f(60\mu m)/f(12\mu m)$ to $f(1.2\mu
m)/f(0.36\mu m)$ flux ratios: (A) prominent big blue bump objects
(e.g. 3C~273), which correspond to the `Normal', E94-like, objects;
(B) rapidly dropping optical SEDs with no strong FIR emission (e.g.
MCG-6-30-15), which are reddened AGN and resemble the `No Big Bump'
objects; (C) strong FIR, weak optical/UV objects (e.g. NGC~3227),
which are dominated by host galaxy emission in both regions and so
correspond to the `Host dominated' objects. The host FIR emission in
Class~C AGN is often extended and lies in the range of normal
galaxies in 50\% of cases, and of starbursts in the rest.

In a companion paper (Carleton et al. 1987) the same authors argued
that all of these types were consistent with a single intrinsic form
of the quasar SED, modified only by obscuration and host galaxy
contamination. The bluest of the Class~A objects would exhibit this
SED form, which is close to the E94 mean SED. Many of the `reddened
Class~B' objects have X-ray column densities
N$_H\sim$10$^{22}$-10$^{23}$~cm$^{-2}$, and have intermediate AGN
types (1.5, 1.8, 1.9 Osterbrock \& Koski 1976) indicating modest
reddening of the broad emission line region and continuum of
A$_V\sim$0.5-3. We do not have H$\alpha$/H$\beta$ ratios, or X-ray
N$_H$ values, for more than a handful of the COSMOS type~1 AGN
sample and so cannot yet test whether the XMM-COSMOS sample shares
these properties with the Piccinotti AGN. We explore the
AGN-host-reddening parameter space in detail in Paper 3.

The fourth class, D. ``hot-dust-poor'', has not been recognized
before. The missing inflection point is apparently due to the
reduced `hot dust' bump by at least a factor 3 in the near-IR. As
the hot dust emission is attributed to the `torus' invoked in
unified models of AGN to explain the type~1/type~2 dichotomy, the
existence of a class of AGN without this `torus' emission raises
questions about the universality of the unified model. This class of
sources are discussed in detail in Hao et al. (2010) for COSMOS and
Hao et al. (2011a) for optically selected samples.

\section{Summary \& Conclusions}

In this paper we have assembled a large sample of 413 type~1 AGN
(emission line FWHM$>2000$~km~s$^{-1}$), selected in X-rays by \xmm\
within the COSMOS field. The sample includes sources with
spectroscopic redshifts and a uniform multiwavelength coverage from
the X-ray to the far infrared, with 33.6\% radio (VLA) coverage too.

We briefly analyzed the optical and X-ray properties of the sample
with the main aim to compare them with the Elvis et al. (1994)
sample, which is our reference for the construction of the mean SED.

The main goal of this paper is to derive the mean Type 1 AGN SED. We
have derived SEDs for all the 413 type~1 AGN. For 203 sources, we
could use the black hole mass and the $M_{BH}-M_{bulge}$ scaling
relation to produce host-subtracted SEDs. The COSMOS type~1 AGN
sample spans a much larger range of redshift and luminosity than the
E94 sample. Only 6 of these AGN, $\sim$1.5\%, are radio-loud.

These SEDs make use of the vast COSMOS photometric data set, and so
contain many photometric points (a mean of 35 per SED). The SEDs are
especially well sampled in the optical/UV band ($\sim$0.1 -- 1
$\mu$m), with a mean of 18 photometric points. We have corrected the
SEDs for Galactic extinction, have restricted the data collection
time interval to limit variability and have made a correction for
the BEL contribution. The SEDs were re-sampled on a uniform rest
frame frequency grid.

The mean SED in the rest frame 8 $\mu m$ to 4000\AA are calculated
based on detections only. Mean SED beyond this range are calculated
based on reasonable power law assumption. The mean SED of the sample
before host galaxy correction is quite different from the E94 mean
radio-quiet SED and does not show a prominent 1$\mu$m inflection
point between the UV and near-IR bumps. Sub-samples of AGN (with
black hole mass estimation) corrected for host galaxy contribution
using the Marconi \& Hunt (2003) scaling relationship, with an
additional redshift evolution term, restores the 1$\mu$m inflection
to the mean COSMOS type~1 SED. This shows that host galaxy
contamination is likely to be a major contributor to the variety of
SEDs in an X-ray selected sample. Evolution of the M-$\sigma$
relation is supported by the better host subtraction obtained using
the Bennert et al. (2011) relation.

The SEDs before and after the corrections for the Galactic
extinction, for broad emission line contributions, constrained
variability, and for host galaxy contribution are available on-line
at the journal web site.

Both host contamination and reddening for all COSMOS quasars will be
addressed in companion papers (Hao et al. 2012 in prep.).

Several extreme types of AGN SED were identified, corresponding to
an E94 SED, a reddened E94 SED and a host dominated SED.  These
three SED types have been seen previously (e.g. Ward et al. 1985). A
new sub-class of `hot-dust-poor' quasars has been found that appears
to lack strong hot dust emission. These will be of interest for
unification models as they appear to indicate the absence of the
standard obscuring `torus'. Their properties are discussed in Hao et
al. (2010, 2011a).

The COSMOS AGN photometry coverage continues to expand. In
particular the far-infrared to millimeter region: Far-IR
(60-600$\mu$m) has been observed with the {\em Herschel} satellite,
SCUBA2 will image the field at 850$\mu$m and 1.1~mm imaging of much
of the field by AzTeC has been performed and should yield several
hundred sources (Scott et al. 2008, Austermann et al. 2009). The
quasar SED properties over wider wavelength ranges will be possible
in the near future.

\section*{Acknowledgments}

ME and HH thank J. McDowell for useful discussion. This work was
supported in part by NASA \chandra grant number GO7-8136A (HH, FC,
ME), NASA contract NAS8-39073 (Chandra X-ray Center) and the
Smithsonian Scholarly Studies (FC). KJ acknowledges support from the
Emmy Noether Programme of the German Science Foundation (DFG)
through grant number JA~1114/3-1. In Italy this work is supported by
ASI/INAF contracts I/009/10/0, I/024/05/0 and I/088/06. In Germany
this project is supported by the Bundesministerium f\"{u}r Bildung
und Forschung/Deutsches Zentrum f\"{u}r Luft und Raumfahrt and the
Max Planck Society. MS and GH acknowledge support bei the Leibniz
Prize of the Deutsche Forschungagemeinschaft, DFG (HA 1850/28-1). YL
acknowledges partial funding support by the Directional Research
Project of the Chinese Academy of Sciences under project no.
KJCX2-YW-T03 and by the National Natural Science Foundation of China
under grant nos. 10821061, 10733010, 10725313, and by 973 Program of
China under grant 2009CB824800.


\end{document}